\documentclass[]{aa}
\usepackage[varg]{txfonts}
\usepackage[usenames]{color}
\usepackage{graphicx}
\usepackage{epsfig}
\usepackage{natbib}
\usepackage{journals_abrv}
\usepackage{amssymb}
\bibpunct{(}{)}{;}{a}{}{,} 

\newcommand{\geff}{\ensuremath{\mathit{g}_{\rm eff}}}
\newcommand{\gline}{\ensuremath{\mathit{g}_{\rm rad}^{\rm line}}}

\newcommand{\mdot}{\ensuremath{\dot{M}}}          

\newcommand{\msun}{\ensuremath{\mathit{M}_{\odot}}}

\newcommand{\lsun}{\ensuremath{\mathit{L}_{\odot}}}
\newcommand{\rstar}{\ensuremath{\mathit{R}_{\star}}}    
\newcommand{\rsun}{\ensuremath{\mathit{R}_{\odot}}}
\newcommand{\teff}{\ensuremath{\mathit{T}_{\rm eff}}}   
\newcommand{\vinf}{\ensuremath{\mathit{\varv}_{\infty}}}    
\newcommand{\vesc}{\ensuremath{\mathit{\varv}_{\rm esc}}}  

\newcommand{\beq}{\begin{equation}} 
\newcommand{\eeq}{\end{equation}}

\begin{document}

\title{Predictions for mass-loss rates and terminal wind velocities of massive O-type stars}

\author{ L. E. Muijres \inst{1}       \and
 	     Jorick S. Vink \inst{2}       \and
            A. de Koter \inst{1,3}      \and
	     P.E. M\"uller \inst{4}        \and
	     N. Langer \inst{5,3}
       }

\institute{ Astronomical Institute 'Anton Pannekoek', University of Amsterdam, 
           Science Park 904, NL-1098 Amsterdam, The Netherlands
	       \and
           Armagh Observatory, College Hill, Armagh BT61 9DG, Northern Ireland, UK
	       \and
	    Astronomical Institute, Utrecht University, Princetonplein 5, 
           3584 CC Utrecht, The Netherlands
	       \and
	    School of Physical and Geographical Sciences, Lennard-Jones Laboratories, Keele University, 
	    Staffordshire, ST5 5BG, United Kingdom
	       \and
           Argelander-Institut f\"{u}r Astronomie der Universit\"{a}t Bonn, 
           Auf dem H\"{u}gel 71, 53121 Bonn, Germany	      
          }

\date{Received / Accepted}

\abstract
{Mass loss from massive stars forms an important aspect of the evolution of massive stars, as well as for 
the enrichment of the surrounding interstellar medium.}
{Our goal is to predict accurate mass-loss rates and terminal wind velocities. 
These quantities can be compared to empirical values, thereby testing
radiation-driven wind models. One specific topical issue is that of the so-called ``weak-wind problem'', where empirically 
derived mass-loss rates and (modified) wind momenta fall orders of magnitude short of predicted values.}
{We employ an established Monte Carlo model and a recently suggested new line acceleration formalism 
to solve the wind dynamics more consistently.}
{We provide a new grid of mass-loss rates and terminal wind velocities of $\textrm{O}$-type stars, and compare 
the values to empirical results. Our models fail to provide mass-loss rates for main-sequence stars 
below a luminosity of log$(L/\lsun) = 5.2$, where we appear to run into a fundamental limit. 
At luminosities below this critical value there is insufficient momentum transferred to the wind in the region below the sonic point in order 
to kick-start the acceleration of the flow. This problem occurs at almost the exact location of 
the onset of the weak-wind problem. For O dwarfs, the boundary between being able to start a wind, and failing to do so, 
is at spectral type O6/O6.5. The direct cause of this failure for O6.5 stars is a combination of the lower luminosity and
a lack of Fe\,{\sc v} lines at the base of the wind.
This might indicate that -- in addition to 
radiation pressure -- another mechanism is required to provide the necessary driving to initiate the wind acceleration.}
{For stars more luminous than $10^{5.2}$\,\lsun, our new mass-loss rates are in excellent 
agreement with the mass-loss prescription by Vink et al. 2000 using our terminal wind velocities as input to this recipe. 
This implies that the main assumption entering the method of the Vink et al. 
prescriptions -- i.e. that the momentum equation is not explicitly solved for -- does not compromise 
the reliability of the Vink et al. results for this part of parameter space. Finally, our new models predict terminal velocities that are typically 
35 and 45 percent larger than observed values. Such over-predictions are similar to those from (modified) CAK-theory.}

\keywords{Stars: early-type, massive, mass-loss, winds, outflows, Radiative transfer}

\titlerunning{Predictions for mass-loss rates and terminal velocities}
\maketitle

\authorrunning{}

\section{Introduction}
\label{sec:Introduction}

In this article, we present predictions for mass-loss rates and velocity structures 
for a grid of $\textrm{O}$-type stars, using two distinct methods for solving the wind 
dynamics. 

Mass loss forms an integral aspect characterizing massive $\textrm{O}$-type stars. 
Because of their short lifetimes, massive stars are important tracers of star formation 
in galaxies. Furthermore, they enrich the interstellar medium with metals, both during 
their lives via stellar winds, as well as when they explode at the very end of their 
evolution. In order to build an evolutionary framework for massive stars, it is essential
to map the mass-loss processes during the various evolutionary stages, as the exact rates of 
mass loss greatly influence the evolutionary tracks \citep[e.g.][]{1981A&A....99...97M, 
1986ARA&A..24..329C}. 
The effects of mass loss on the evolutionary tracks are at least two-fold: first and 
foremost the stellar mass is reduced, and secondly, the rotational velocity is strongly 
affected, as the mass also carries away angular momentum \citep[e.g.][]{1998A&A...329..551L,2000A&A...361..159M}. 

For the continuous stellar winds from massive stars, the outflow is thought to be driven 
by the transfer of energy and momentum from the radiation field to the atmosphere through 
the absorption of photons in atomic transitions. The exact amount of momentum 
and energy transfer has been the subject of both theoretical and observational studies for 
many decades ~\citep{1970ApJ...159..879L,1975ApJ...195..157C,1986A&A...164...86P,
1996A&A...305..171P,1997ApJ...477..792D,1999A&A...350..181V,2004A&A...417.1003K,2007A&A...473..603M}.  
For luminous $\textrm{O}$-type stars, with log($L/\lsun) > 5.2$, the 
theoretical predictions of \cite{2000A&A...362..295V} seem to be in reasonable agreement 
with empirical mass-loss rates provided that $\textrm{O}$-stars are only subject to modest amounts of wind 
clumping (with clump filling factors of only 5-10). However, for objects with luminosities 
log($L/\lsun)$ below approximately 5.2, a severe drop -- by a factor of $\sim$100 -- in 
the empirically determined modified wind momentum (basically a multiplication 
of the mass-loss rate and the terminal velocity) has been revealed.
This problem has in literature been referred to as ``the weak-wind problem'' 
~\citep{1996A&A...305..171P,2005A&A...441..735M,2008A&ARv..16..209P,2009A&A...498..837M}.  

It deserves proper investigation simply because of the enormity of the effect.
It is particularly important to find out whether the problem is caused by the mass-loss 
diagnostics or the predictions, as both are also applied to more luminous stars, where 
agreement between diagnostics and predictions has seemingly been achieved. But how 
certain can we be that this agreement is not a coincidence if we are aware of severe problems at lower luminosity? 

Furthermore, we note that the oft-used mass-loss predictions of \cite{2000A&A...362..295V} are  
semi-empirical, in the sense that empirical values for the wind velocity structure and terminal 
velocity are used as input to the modelling. 
In order to trust our overall knowledge of the mass-loss rates from $\textrm{O}$-type 
stars -- at {\it all} masses and luminosities -- it is pivotal to 
further scrutinize the \cite{2000A&A...362..295V} assumptions, most notably that of the adopted 
wind dynamics.

Recently, \cite{2008A&A...492..493M} suggested a new parametrization of the line acceleration, 
expressing it as a function of radius rather than of the velocity 
gradient, as in \citeauthor{1975ApJ...195..157C} 
(\citeyear{1975ApJ...195..157C}; henceforth CAK) theory. The implementation of this new formalism allows 
for local dynamical consistency, as one can determine the energy and momentum transfer 
at each location in the wind through the use of Monte Carlo simulations.  
Although the formalism was applied with three independent 
starting conditions that showed convergence to the same wind parameters, it
has thus far only been applied to one object, that of an O5 dwarf.

For the adopted line force parameterization \citeauthor{2008A&A...492..493M} identify an 
exact solution in case the medium is isothermal. 
We expand on this result by also accounting for a temperature stratification. To allow for such
a study we employ the new
line acceleration parameterization but solve for the wind dynamics consistently by applying a
numerical  method to solve for the momentum equation.

The purpose of our study is threefold: {\em    i)} to solve the wind dynamics 
numerically, and compare the results to those of \cite{2008A&A...492..493M}, {\em    ii)} to 
compute a larger grid of dynamically derived O-star mass-loss rates and wind 
terminal velocities, and determine the accuracy of the predictions
made by \cite{2000A&A...362..295V}, and {\em    iii)} 
to utilize the grid in order to investigate the weak-wind problem.

Our paper is organized as follows. In Sect.~\ref{sec:Method}, we start off 
describing the core of our method and the different methods to treat the wind equation. 
The results are presented in Sect.~\ref{sec:Results} and discussed in the 
Sect.~\ref{sec:discussion}. We end with the conclusions (Sect.~\ref{sec:Conclusions}).

\section{Method}
\label{sec:Method}

The method of \citet{1997ApJ...477..792D} and \citet{1999A&A...350..181V}, applied to derive the mass-loss rates
of O and early-B type stars \citep{2000A&A...362..295V,2001A&A...369..574V}, 
Luminous Blue Variable stars \citep{2002A&A...393..543V} and Wolf-Rayet stars 
\citep{2005A&A...442..587V}, is an extension of a treatment developed by
\citet{1985ApJ...288..679A}. It is based on an iteration cycle between the
stellar atmosphere model {\sc isa-wind} \citep{1993A&A...277..561D} and a
Monte Carlo simulation, {\sc mc-wind} \citep{1997ApJ...477..792D}, in which the energy per unit time $\Delta L$ 
that is extracted from the radiation field in interactions of photons with the gas, is 
computed.
From this a mass-loss rate \mdot\ is computed on the basis of
which a new {\sc isa-wind} model is constructed. The predicted mass loss is 
the one for which the input mass-loss rate of {\sc isa-wind} equals the 
mass-loss rate computed by {\sc mc-wind}.
  
As is consistently pointed out in the papers referred to above, the method
does not address the equation of motion but uses a prescribed trans-sonic velocity 
structure. This implies that although in a global sense the method fulfills 
the constraint of energy conservation, it need not hold that the actual
local forces acting on the gas are consistent with the force implied by the
adopted velocity law. \cite{2008A&A...492..493M} relax on this assumption
and present an improved treatment of the problem introducing a new way
to parametrize the line force. 
We first discuss an approach presented by these authors, 
which we refer to as the ``best-$\beta$ method, as it allows to link to empirically derived estimates of the steepness of the velocity law 
(characterized by a parameter $\beta$, see below).
In a second step, we present solutions that numerically solve the wind dynamics.

We first briefly introduce {\sc isa-wind} in Sect.~\ref{sec:isawind}, 
emphasizing the treatment of the heuristic velocity law, and {\sc mc-wind} 
in Sect.~\ref{sec:mcwind}, focusing on the determination of the mass-loss 
rate using the global energy argument. In Sect.~\ref{sec:lineforce} we recapitulate 
the essentials of the parametrization of the line force by \citeauthor{2008A&A...492..493M}
and in Sect.~\ref{sec:bestbeta} the principle of their best-$\beta$ method. In the following subsection 
we introduce our hydrodynamical method. Finally, Sect.~\ref{sec:sonicpoint1} is 
devoted to a discussion of the physical conditions at the sonic point.

\subsection{The model atmosphere}
\label{sec:isawind} 

The code {\sc isa-wind} computes the structure, radiation field and
ionization/excitation state of the gas of an outflowing stellar 
atmosphere in non local thermodynamic equilibrium (non-LTE), assuming
radiative equilibrium. No artificial separation between the photosphere
and wind is assumed.
The temperature structure is treated somewhat simplified in
that it results from initial LTE based Rosseland opacities (i.e. grey).
The fact that the temperature structure is not affected by possible 
departures from the populations from their LTE state implies that the
effect of line blanketing is not treated self-consistently, although 
non-LTE line blocking is taken into account. 
Radiation transfer in spectral lines is treated in the Sobolev 
approximation~\citep{1960mes..book.....S}. 

The input stellar parameters are the luminosity $L$, the effective temperature \teff\
(specifying the radius $R$),
the mass $M$ and chemical abundances. The wind is described by the mass-loss
rate \mdot\ and a velocity structure, which are connected through the equation
of mass continuity 
\beq
\label{eq:continuity}
\dot{M} = 4 \pi r^2 \varv(r) \rho(r),
\eeq 
where $\rho(r)$ is the mass density and $v(r)$ is the velocity at radius $r$. 
Outside the photosphere, the velocity structure is assumed to follow a
$\beta$-law, i.e.
\beq
\label{eq:betalaw}
\varv(r) = \vinf \left( 1- \frac{r'}{r}\right)^{\beta}.
\eeq 
The free parameter $\beta$ is a measure of the velocity gradient. A low value of $\beta$
implies that the velocity approaches the terminal flow velocity \vinf\ relatively 
close to the star; for a large value this happens only further out in the wind.
The $\beta$-law does not hold in the photosphere since the line force is not the 
dominant term in the equation of motion, but gravity and the acceleration due
to the gas pressure gradient also contribute to the flow 
structure. The radius $r'$ is a smoothing 
parameter that is used to connect the $\beta$-law to the (quasi) hydrostatic 
photosphere and must assure that $\varv(r)$
and its spatial derivative are continuous at the point where one couples the
photospheric velocity law to the $\beta$-law.
The velocity structure in the photosphere is determined by solving
the non-isothermal equation of motion, neglecting line radiation pressure and
assuming that continuum radiation pressure is the result from Thomson 
scattering only. An inner boundary velocity (or density) is chosen, which may 
be used to tune the total Rosseland optical depth of the photosphere and
wind (see below).

The wind is assumed to be homogeneous, i.e. the outflowing gas is not clumped
\citep[but see][]{2011A&A...526A..32M}, and 
the terminal velocity is chosen to be $2.6$ times the effective escape velocity 
from the stellar photosphere, which is in reasonable concordance with empirically 
determined terminal velocities 
of O-stars ~\citep{1995ApJ...455..269L,2000ARA&A..38..613K}.
The base of the photosphere is positioned at a Rosseland optical depth of about
20-25 and the wind extends out to 20 \rstar.

\subsection{The Monte Carlo method {\sc mc-wind}}
\label{sec:mcwind}

The code {\sc mc-wind} uses the model atmosphere structure computed by {\sc isa-wind}
to determine the total amount of energy that is transferred
from the radiation field to the wind -- in interactions of photons with ions in the
gas -- by means of a Monte Carlo simulation of the trajectories of photon packets 
emitted at the base of the photosphere and escaping through the outer boundary
of the model.
Each photon can travel an optical depth weighted (random) distance to a 
point of interaction. This point is determined by taking into account all the 
opacity the photon encounters on its path, so it includes contributions from 
both lines and continua. At the point of interaction the type of interaction
is determined, using proper weighing functions \citep{1999A&A...350..181V}. 
The possible interactions are
thermal absorption and emission, electron scattering and line scattering.
The interaction is assumed to be coherent in the 
co-moving frame of the ion. In the observers frame, however, energy can be
exchanged from the radiation field to the gas (or vice versa). It is
traced which ion is involved in the interaction, such that, for instance, the 
contributions to the radiative force can be dissected and identified. 
This provides us with a powerful tool to study the nature of the line force at 
each location in the wind.

The radiative force per unit mass equals  ~\citep{1985ApJ...288..679A}:
\beq
\label{eq:mclineforce}
\mathit{g}_{\rm rad} = -\frac{1}{\mdot} \frac{d L}{d r}, 
\eeq     
where $dL$ is the amount of energy lost by the 
radiation field in a layer of thickness $dr$.

Once the total amount of energy transferred to the wind is known, the mass-loss rate that 
can be driven {\em for the density and velocity structure of the adopted {\sc isa-wind} model} can be calculated. 
Neglecting enthalpy:
\beq
\label{eq:mcdeltal}
\Delta L = \frac{1}{2} \mdot \left(\vinf^2 + \varv_{\rm esc,N}^2\right),
\eeq     
where $\Delta L$ is the total amount of energy lost by the radiation field and
\beq
\label{eq:vesc}
\varv_{\rm esc,N} = \sqrt{\frac{2GM_*}{R_*}},
\eeq  
is the Newtonian escape velocity from the stellar surface.    
$G$ is the gravitational constant. 
A new {\sc isa-wind} atmosphere, adopting the mass-loss rate as determined in 
{\sc mc-wind}, is computed followed by a new Monte Carlo simulation.
This procedure is repeated until the input mass-loss rate of {\sc mc-wind}
equals the output mass-loss rate. Although the mass-loss rate that is predicted in this
way reflects that in a {\em global} sense the energy that is needed to drive the wind is indeed
extracted from the radiation field, it does not mean that the input line force
(implied by the velocity law) equals the output line force from the Monte Carlo
simulation {\em locally}, i.e. the equation of motion of the wind is not
solved.

Here we improve on this situation using two methods. Both methods A and B 
require a parametrization of the line force predicted by {\sc mc-wind}.
We therefore first discuss this aspect.

\subsection{Line force parametrization}
\label{sec:lineforce}

\begin{figure}
\begin{center}
     \includegraphics[width=0.45\textwidth,angle=0]{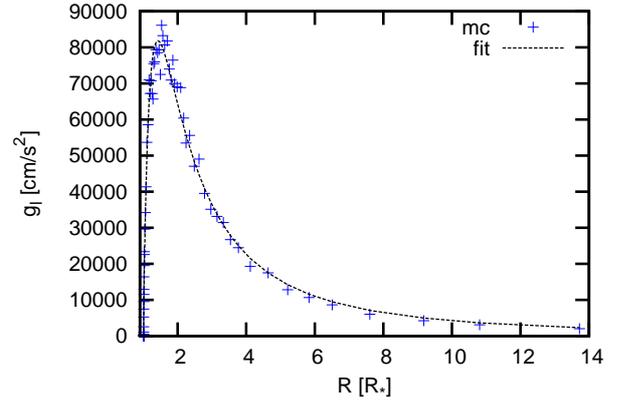}
     \caption{The line force (blue crosses) as predicted by {\sc mc-wind} in the first iteration
              step. A fit (black dotted line) using Eq.~\ref{eq:grad} to represent the
              line force is overplotted. Note the modest scatter on the Monte Carlo results due to
              noise.}
  \label{fig:mcline}
  \end{center}
\end{figure}

Figure~\ref{fig:mcline} shows the Monte-Carlo line force (blue crosses) as 
is produced in the first iteration step of a typical O3\,V star 
($L = 10^{5.83}$ \lsun, \teff = 44,600\,K and $M = 58\,\msun$). 
The Monte Carlo line force is determined in a statistical way and
shows scatter. Given the delicate nature of the equation of motion it
can not be used as such and must be represented by an appropriate 
analytical fit function. 
We adopt a parametrization of the line force as a function 
of radius, rather than of optical depth, as opted for by \citet{1975ApJ...195..157C}. 
In Sect.~\ref{sec:MCAK-theory} we show that this leads to a more accurate numerical 
representation of the line force, at least for the type of stars studied here.
In doing so, we follow 
\cite{2008A&A...492..493M}, who motivate
\beq
\label{eq:grad}
\gline = \left\{ \begin{array}{rl} 
       0 & \hspace{6mm} \textrm{if  } \hspace{2mm} r < r_{\circ} \\
       \mathit{g}_{\circ} \,(1-r_{\circ}/r)^{\gamma} / r^{2} & 
       \hspace{6mm} \textrm{if  } \hspace{2mm} r \geq r_{\circ}, \\
	              \end{array} \right.		
\eeq
where $\mathit{g}_{\circ}$, $r_{\circ}$ , and $\gamma$ are fit parameters to the Monte Carlo line force. 
This choice of the fit function, i.e. without any explicit dependency of the line force on the velocity gradient, implies that in our
models the critical point is the sonic point.
Figure~\ref{fig:mcline} shows a typical result for this fit (black 
dotted curve). The deviations are (as mentioned) due to scatter in the 
simulation.

\subsection{Method A: Best-$\beta$ solution}
\label{sec:bestbeta}

In this section, we use the line force representation Eq.~\ref{eq:grad} to determine --
after making certain assumptions -- an analytical solution of the velocity law in the 
outer part of the wind, following \cite{2008A&A...492..493M}. This solution can be compared 
to the $\beta$-law (Eq.~\ref{eq:betalaw}) and used to derive \vinf\ and the  
$\beta$ value that is most representative. This is useful in comparing to the often 
applied $\beta$-law.

We aim to find a solution of the equation of motion 
\beq
\label{eq:motion}
v \frac{dv}{dr} = -\frac{R_* \vesc^2}{2 r^2} + \gline - \frac{1}{\rho}\frac{d p}{d r}, 
\eeq
where $p$ is the gas pressure and 
\beq
\label{eq:vesceff}
\vesc = v_{\rm esc,N} \sqrt{1-\Gamma}, 
\eeq
is the effective surface escape velocity of the star. $\Gamma$ is the continuum radiation
pressure in units of the Newtonian gravitational acceleration. Sufficiently far from the
photosphere this term is dominated by radiation pressure on free electrons, i.e.
$\Gamma$\,=\,$\Gamma_{\rm e}$, where $\Gamma_{\rm e}$ is essentially constant for
early-type stars. 
Close to or in the photosphere, the acceleration due to free-free, 
bound-free and bound-bound processes may compete 
with electron scattering and should, in principle, be considered in
Eq.~\ref{eq:vesceff}. For our best-$\beta$ solution, however, we assume a constant continuum
acceleration, which we set to $\Gamma_{\rm e}$.
Substituting the equation of state for an ideal gas and using Eq.~\ref{eq:continuity}, 
the term $(1/\rho)\,dp/dr$ can be written as:
\beq
\label{eq:pressure}
\frac{1}{\rho}\frac{d p}{d r} = -\frac{a^2}{v} \frac{dv}{dr} - \frac{2 a^2}{r} + \frac{k}{m} \frac{d T}{d r},
\eeq
where $k$ is Boltzmann's constant, $m$ the mean particle mass and $a(r)$ is the local sound 
speed, given by:
\beq
a  = \sqrt{\frac{k T}{m}}.
\eeq
We assume the wind to be isothermal, such that the sound speed is constant.
The equation of motion can now be rewritten as
\beq
\label{eq:motionfinal}
a_{\circ}  \left(\frac{v}{a_{\circ}} - \frac{a_{\circ}}{v} \right)\frac{dv}{dr} = -\frac{R_* \vesc^2}{2 r^2} + \frac{2 a_{\circ}^2}{r} + \gline, 
\eeq
where $a_{\circ}$ is the isothermal sound speed at the effective temperature of the star.
Equation \ref{eq:motionfinal} is a critical point equation, where the left- and right-hand 
side vanish at the point $v(r_s)=a_{\circ}$, i.e. where $r_s$ is the radius of the sonic point. It yields several types of solutions.
\citet{2008A&A...492..493M} show that for the isothermal case and a line force as described in Eq.~\ref{eq:grad}, analytical expressions for all types
of solutions of Eq.~\ref{eq:motionfinal} can be constructed by means of the Lambert W function (as for a further discussion of the solution of
Eq.~\ref{eq:motionfinal} containing an additional centrifugal term, see
\citeauthor{2001mueller} \citeyear{2001mueller}). Even for the interesting trans-sonic case of a stellar wind, the 
analytical solution has an intricate shape. However, a useful approximate wind solution for the velocity law 
can be constructed if the pressure related terms $2a^{2}/r$ and
$a/v$ can be neglected. We note, however, that 
at the sonic point the contribution of the two pressure terms is non-negligible \citep{2008A&A...492..493M}. 
After some manipulation one finds that the approximate velocity law is given by:
\beq
\label{eq:vlawapprox}
v(r) = \sqrt{ \frac{R_* v_{\rm esc}^2}{r} +  \frac{2}{r_{\circ}}\frac{\mathit{g}_{\circ}}{\left( 1+\gamma \right)} \left(1-\frac{r_{\circ}}{r} \right)^{\gamma + 1} + C},
\eeq
where $C$ is an integration constant. From this equation, the terminal wind 
velocity can be derived if the integration constant $C$ can be determined. 
This can be done assuming that at radius $r_{\circ}$ the velocity approaches zero.
This yields
\beq
\label{eq:integrationc}
C = - \frac{R_* \vesc^2}{r_{\circ}}.
\eeq
In the limit $r \rightarrow \infty$ we find that:
\beq
\label{eq:vinf}
\vinf = \sqrt{\frac{2}{r_{\circ}} {\frac{\mathit{g}_{\circ}}{(1+\gamma)} - \frac{R_* \vesc^2}{2}}}.
\eeq

The terminal velocity
$\vinf$ can also be determined from the equation of motion. At the critical point, 
the left-hand and right-hand side of Eq.~\ref{eq:motionfinal} both equal zero. Introducing
\vinf\ in relation to $\mathit{g}_{\circ}$ as expressed in Eq.~\ref{eq:vinf}, we find
\beq
\label{eq:vinfnew}
v_{\infty,{\rm new}} = \sqrt{\frac{2}{r_{\circ}} \left[ \left( \frac{r_s}{r_s-r_{\circ}} \right)^{\gamma} 
                 \frac{r_s}{(1+\gamma)} \left( \frac{\vesc}{2} - 2 r_s \right) - \vesc^2 \right]}.
\eeq

A direct comparison to the $\beta$-law can be made for the 
supersonic regime of the wind and results in 
\beq
\label{eq:beta}
\beta = \frac{1+\gamma}{2}.
\eeq
Given the assumptions made in this derivation, this result is only approximately 
correct. 

The procedure that is followed to obtain the best-$\beta$ solution is that
in each Monte Carlo simulation the values of $\mathit{g}_{\circ}$, $r_{\circ}$, and $\gamma$ are 
determined by fitting the output line force. Using these values and the current value of
the sonic point radius, Eqs.~\ref{eq:vinf},~\ref{eq:vinfnew} and~\ref{eq:beta} are used to determine
$\vinf$ and $\beta$. \vinf\ derived from Eq.~\ref{eq:vinfnew}, the mass loss predicted
in {\sc mc-wind}, and the expression derived for $\beta$ serve as input for a new {\sc isa-wind} model. 
The two codes are iterated until convergence is achieved.

Following \cite{2008A&A...492..493M}, we assume that convergence is achieved 
when the values for \vinf\ derived from Eqs.~\ref{eq:vinf} and~\ref{eq:vinfnew} 
agree within 10 percent. This implies that our predicted terminal velocities have at least
this uncertainty. Once the velocity convergence criterion is fulfilled 
all fit parameters and the values for \mdot\ and the sonic point radius will be stable to 
within five to ten percent.

\subsection{Method B: Hydrodynamic solution}
\label{sec:ns}

The accuracy of the best-$\beta$ solution hinges on the assumptions that the wind is isothermal
and that the Eddington-factor $\Gamma$ is constant (here taken to be equal to $\Gamma_{\rm e}$).
It may be expected that these assumptions have an
impact on the velocity structure near the sonic point, which is where the mass-loss
rate is set. To assess this impact and to improve on the physical treatment, we
devise a numerical solution of the equation of motion (Eq.~\ref{eq:motionfinal}) 
throughout the entire photosphere and wind, referred to as the hydrodynamic solution.

To this end we start our solution at the critical point $\varv = a$ and proceed
both down-stream and up-stream using a $4^{\rm th}$ order Runge Kutta method with 
adaptive stepsize control \citep{1992nrfa.book.....P}. Applying l'H\^{o}pital's rule
\citep[see e.g.][]{1999isw..book.....L}
an expression can be devised to determine $dv/dr$ at $v(r_{\rm s}) = a$. In order
to determine the location of the sonic point $r_{\rm s}$ we require
\beq
\label{eq:criticalp}
-\frac{R_* \vesc^2}{2 r_{\rm s}^2} + \frac{2 a^2}{r_{\rm s}} + \gline = 0.
\eeq 
The above equation is solved numerically. 

\begin{figure*}
     \centering
       \resizebox{18.0cm}{!}{
     \includegraphics[width=0.70\textwidth,angle=0]{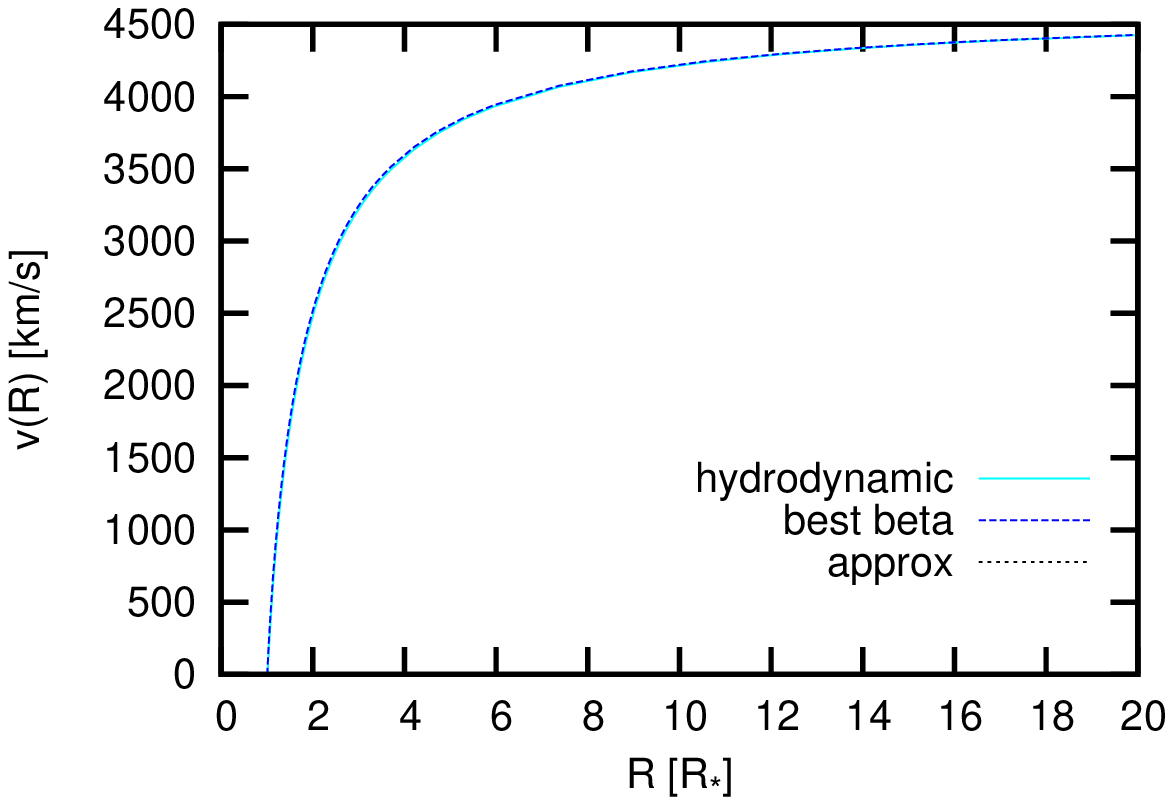}
     \includegraphics[width=0.70\textwidth,angle=0]{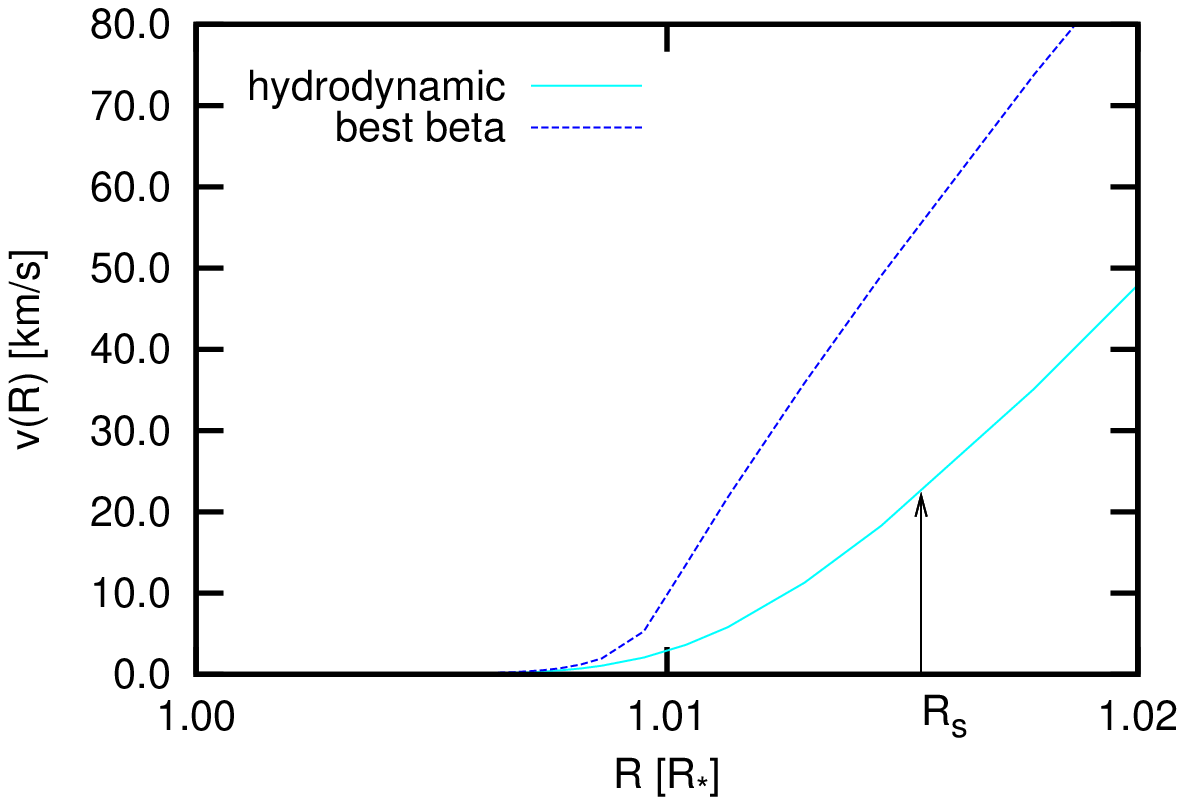}
     \includegraphics[width=0.70\textwidth,angle=0]{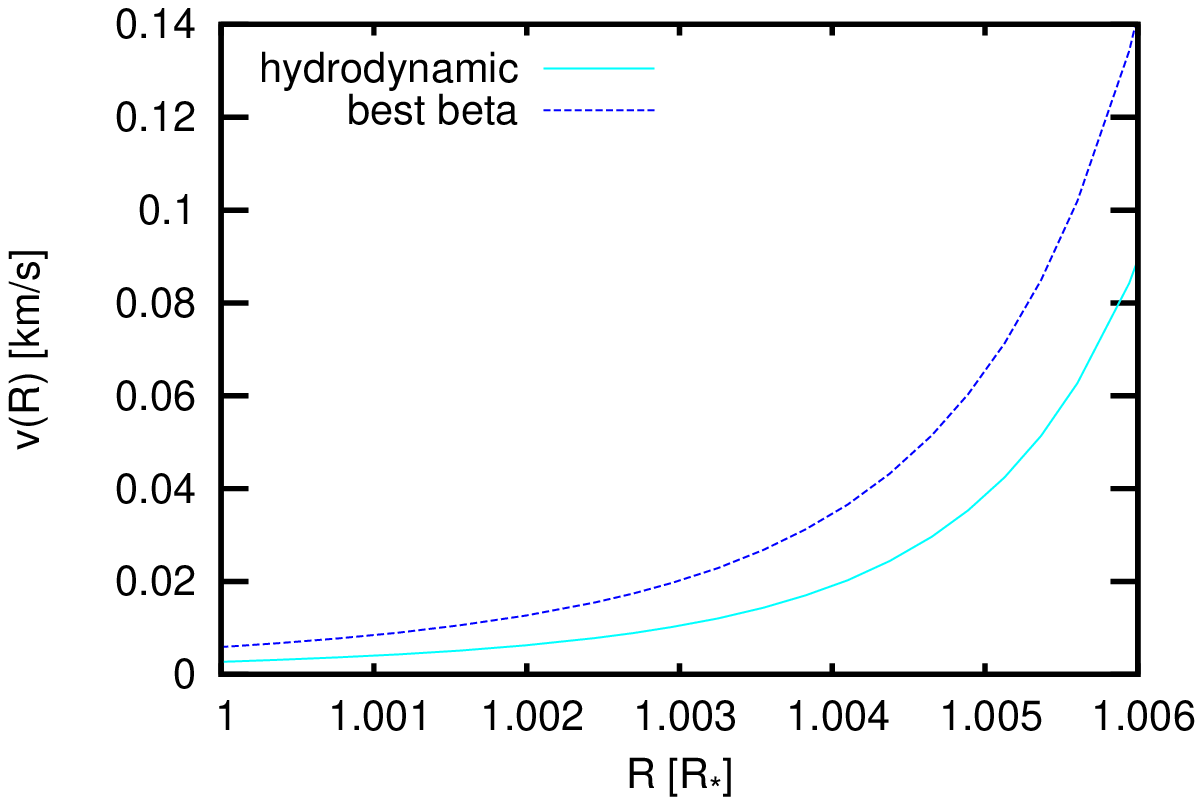}
      }
      \caption{A comparison between the best-$\beta$ and hydrodynamic velocity laws for an O3\,V star. 
          Here, the best-$\beta$ velocity law is the one determined by using the fit-parameters of
          the hydrodynamic solution. The label `approx' in the left panel refers to the approximate velocity law, as given 
	  by Eq.~\ref{eq:vlawapprox}.
	  Three different radial regimes are plotted: full radial range (left panel); the region around the sonic point (middle 
	  panel) in which the location of the sonic point in the hydrodynamic solution is indicated with an arrow, and the
          photospheric region (right panel). Note that at high velocities both methods, as well as the approximate velocity law,
	  yield very similar velocity stratifications. Near the sonic point the best-$\beta$ velocity law is steeper
          than the hydrodynamic velocity law.
          Its sonic point is closer to the star. Given this difference, the right panel shows that in the photosphere where hydrostatic equilibrium controls 
          the equation of motion, the shape of both velocity profiles is very similar. 
          }
  \label{fig:velocitylaw}
\end{figure*}

So far, the hydrodynamic solution assumes an iso-thermal medium. At the sonic point the
temperature gradient is very small, therefore the location of $r_{\rm s}$ can be
reliably determined using Eq.~\ref{eq:criticalp}, even if $dT/dr$ would be taken into account.
The neglect of the temperature gradient in the hydrodynamic solution in the region below 
the sonic point has a significant impact on the structure -- for instance on the
total (Rosseland) optical depth from the inner boundary to the sonic point. 
To solve this problem, 
we account for the temperature structure inward of the critical point.
This requires an iterative procedure
between the solution of the non-isothermal equation of motion
\beq
\label{eq:motionnoniso}
a \left(\frac{v}{a} - \frac{a}{v} \right)\frac{dv}{dr} = -\frac{R_* v_{\rm esc}^2}{2 r^2} + \frac{2 a^2}{r} + \gline - \frac{k}{m} \frac{dT}{dr},
\eeq
and the temperature structure (see Sect.~\ref{sec:isawind}).
After starting numerical integration of the velocity structure at the sonic point $r_{\rm s}$ now determined by
applying Eq.~\ref{eq:criticalp}, but using the 
local value of the temperature at the sonic point) in the down-stream direction, 
we include the $dT/dr$ term
in Eq.~\ref{eq:motionnoniso}. This implies that the location of the sonic point is not affected. 
In the up-stream direction the temperature gradient is negligible, and is ignored.

Figure~\ref{fig:velocitylaw} compares the best-$\beta$ and hydrodynamic velocity laws for an O3 main sequence star.
It shows that the best-$\beta$ solution behaves very similar to the numerical velocity law. 
Note that if one zooms in on the location of the sonic point, one sees that for the best-$\beta$ 
velocity structure $r_{\rm s}$ is positioned slightly more inward, or, alternatively, 
that the velocity law is steeper in the lower part of the wind. In the best-$\beta$ method,
the absolute scaling of the velocity structure in the photosphere is based
on the adopted velocity at the inner boundary of the model (see Sect.~\ref{sec:isawind}).
Therefore, only the position of $r_{\rm s}$ as predicted by the hydrodynamical method is
physically meaningful, albeit in the context of the assumption $\Gamma = \Gamma_{\rm e}$.

Once this iterative procedure has converged, and the non-LTE state of the gas is computed
throughout the atmosphere, we iterate between {\sc isa-wind} and {\sc mc-wind} in the same
manner as described in Sect.~\ref{sec:bestbeta}.
Again, save for \vinf, the fit parameters converge on an accuracy of better than ten percent in
a few iteration cycles. For \vinf\ we are forced to adopt an accuracy of 20 percent. Our
predicted terminal velocities have at least this uncertainty.

\subsubsection{Remaining assumptions and uncertainties}

In the hydrodynamic solution the contribution of bound-free and free-free opacity to the continuum radiation
pressure is ignored (see Sect.~\ref{sec:isawind}). In the photosphere, the contribution of these
processes to $\Gamma$ is not negligible and may in fact be of the order of $\Gamma_{\rm e}$.

We use the Sobolev approximation for line radiation transfer. 
The Sobolev approximation becomes ill-founded for 
small velocity gradients $dv/dr$ or velocities lower than the sound speed. 
\cite{1986A&A...164...86P} showed that in the photosphere (where the velocity is very small) the 
line force is underestimated in the Sobolev approximation.
Further out, in the region of the sonic point, the line optical depth is 
overestimated compared to co-moving frame values, implying an overestimate of the line force
in this region and therefore an overestimate of the mass-loss rate.

In addition to the above two sources of uncertainty to the balance of forces at and below the 
sonic point
is the quality of the fitting function 
Eq.~\ref{eq:grad} in this part of the wind, that may be uncertain by up to a factor of
two. This is not expected to be a big problem deep in the photosphere, as both the fit function as well as the simulated Monte Carlo
line force are small compared to the radiative force on free electrons, but at the sonic point it might play a role.

\subsection{The line force at the sonic point: a test for the validity of the best-$\beta$ method}
\label{sec:sonicpoint1}

The critical point of the equation of motion is the sonic point. A dissimilarity between the sonic point and
the critical point may occur when the line force is represented by an explicit function of $dv/dr$, such as
in CAK and modified-CAK theory~\citep{1986A&A...164...86P}. Although these
descriptions provide extremely valuable insights, they do make assumptions regarding the
behavior of \gline\ (See Sect.~\ref{sec:MCAK-theory}). The same applies for our method. Here we want to
point out that Eq.~\ref{eq:criticalp} implies that -- whatever the description of the line force -- at the sonic
point $\gline \simeq \geff$, as the pressure gradient terms $2a^{2}/r$ and $(k/m)\,dT/dr$ are small 
compared to the line
force. 
This is characteristic for monotonic winds of 
early-type stars \citep[see][for non-monotonic flows]{2000ApJ...532L.125F}. 
Here $\geff = G M_{*} (1-\Gamma) / r^{2} = R_{*} \vesc^{2} / 2r^{2}$. This implies that
for the velocity structure to be a physical solution it must be that at the sonic point
$\gline/\geff \simeq 1$, as pointed out by~e.g. \cite{1975ApJ...195..157C}. We require from our best-$\beta$
solutions, that this criterion is fulfilled. If $\gline/\mathit{g}_{\rm eff}$ is not approximately equal to $1$   
at the sonic point, we interpret this as a failure of the wind to become trans-sonic due to a lack of line force at the
location in the wind where it is essential. Dynamical effects might occur, such as fall back, that are beyond the
topic of this paper. In any case, we interpret such solutions as cases in which the wind cannot be
initiated by line driving alone. For the hydrodynamical solution a failure to fulfill the above requirement 
implies that we do not find a solution at all.

\section{Results}
\label{sec:Results}
\subsection{Grid}
\label{sec:grid}

In order to study our predictions of the wind properties of O-type stars in a systematic manner, we define 
a grid of main sequence, giant and supergiant stars using the 
spectral calibration of \citet{2005A&A...436.1049M} adopting theoretical effective temperature scales. This calibration is based on
non-LTE models that take into account line blanketing effects and an outflowing
stellar wind. We have applied solar abundances as derived by~\cite{1989GeCoA..53..197A},
consistent with the predictions of \cite{2000A&A...362..295V}. For all stars in
the grid, we have derived the mass-loss rate, terminal velocity and $\beta$-parameter.
The hydrodynamic solution does not feature a $\beta$, rather $\gamma$ is the parameter that
describes the slope of the velocity law. To better facilitate a comparison between the different methods
we have applied Eq.~\ref{eq:beta} to convert $\gamma$ into a $\beta$-value, referred to as
$\beta_{\gamma}$. The calculated grid is given in Table~\ref{table:grid-results}.
The final column lists the mass-loss rate as predicted using the fitting
formula of  \cite{2000A&A...362..295V} that assume a fixed value of $\beta = 1$, but 
for an input variable value of $\vinf = 2.6\,\vesc$.

\begin{figure*}
  \centering 
     \resizebox{10.3cm}{!}{
     \includegraphics[width=0.60\textwidth,angle=0]{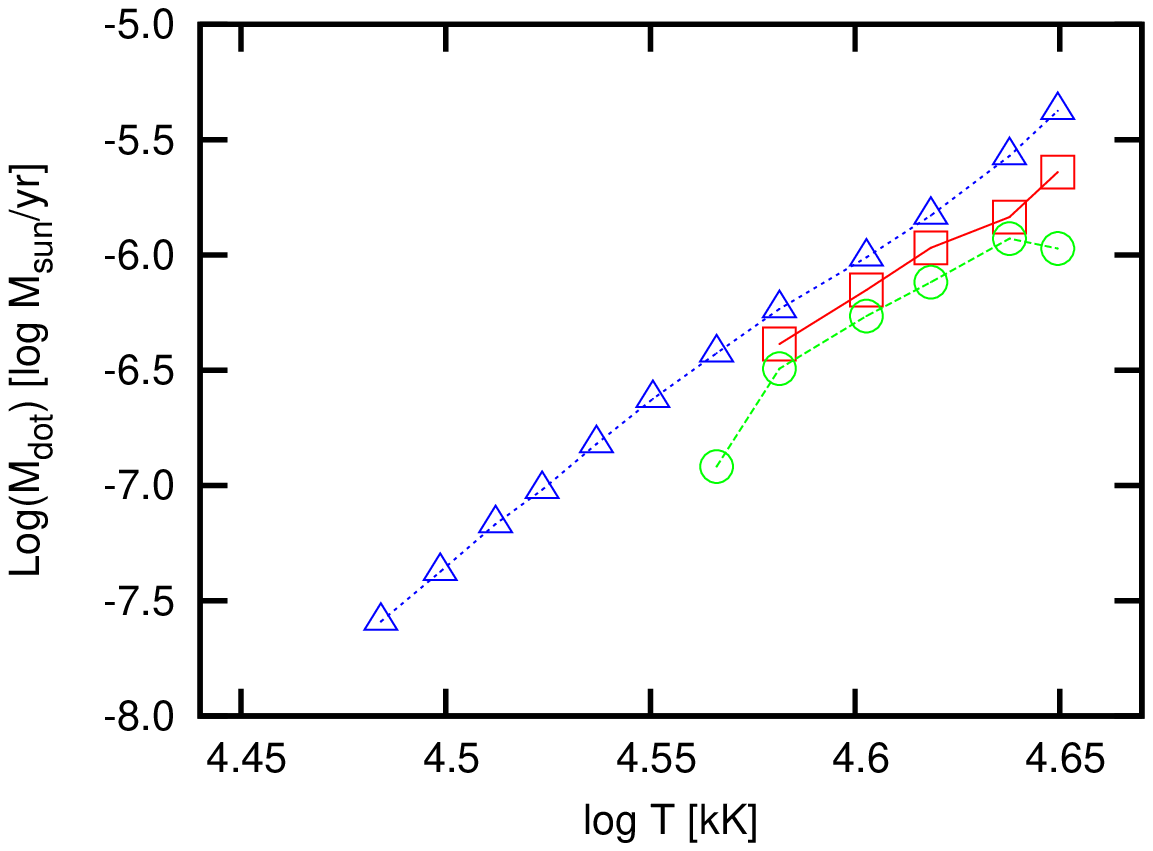}
     \includegraphics[width=0.60\textwidth,angle=0]{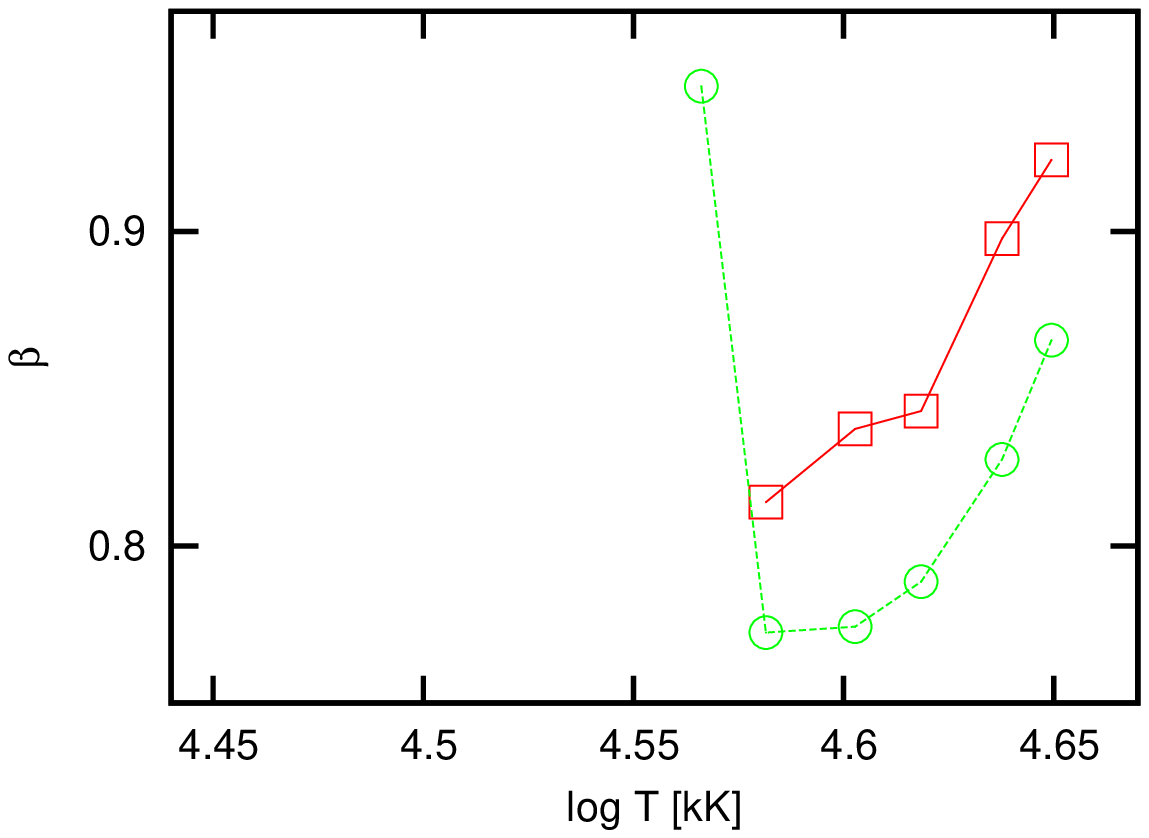} }
     
     \resizebox{10.3cm}{!}{
     \includegraphics[width=0.60\textwidth,angle=0]{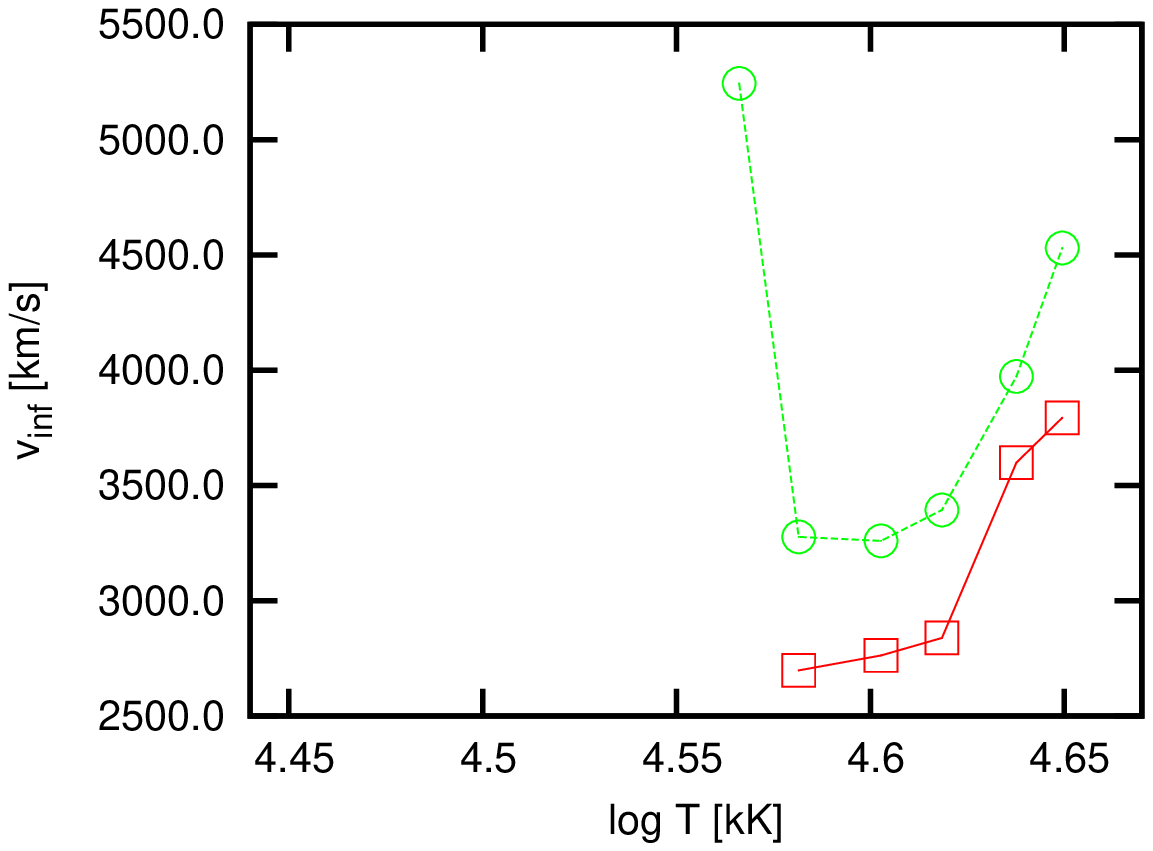}  
     \includegraphics[width=0.60\textwidth,angle=0]{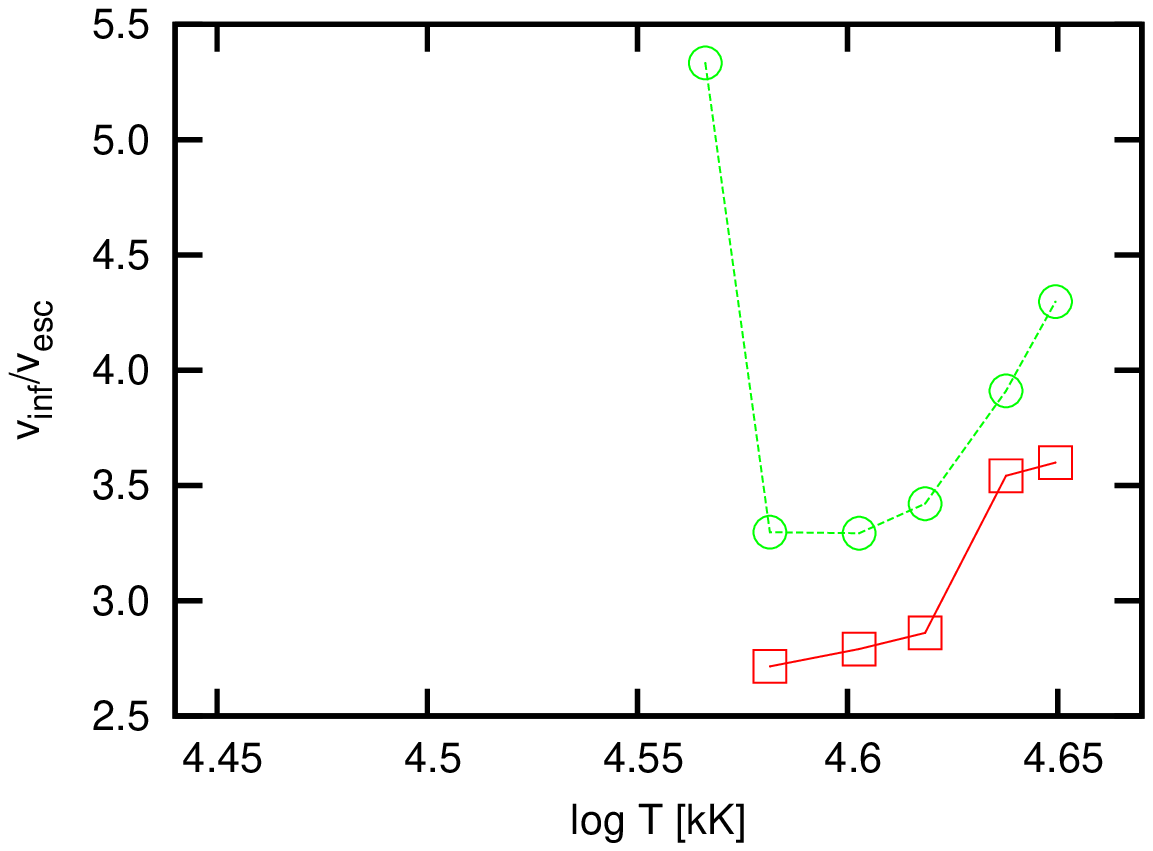} }    
  \caption{Predicted \mdot, \vinf ~and~ $\beta$ for the main sequence stars. Best-$\beta$ solutions are given in red and
           hydrodynamic solutions in green. For comparison, theoretical results by \citet{2001A&A...369..574V} are provided
           in blue.}
  \label{fig:masslossLV}
\end{figure*}

\begin{figure*}
  \centering 
     \resizebox{10.3cm}{!}{
     \includegraphics[width=0.60\textwidth,angle=0]{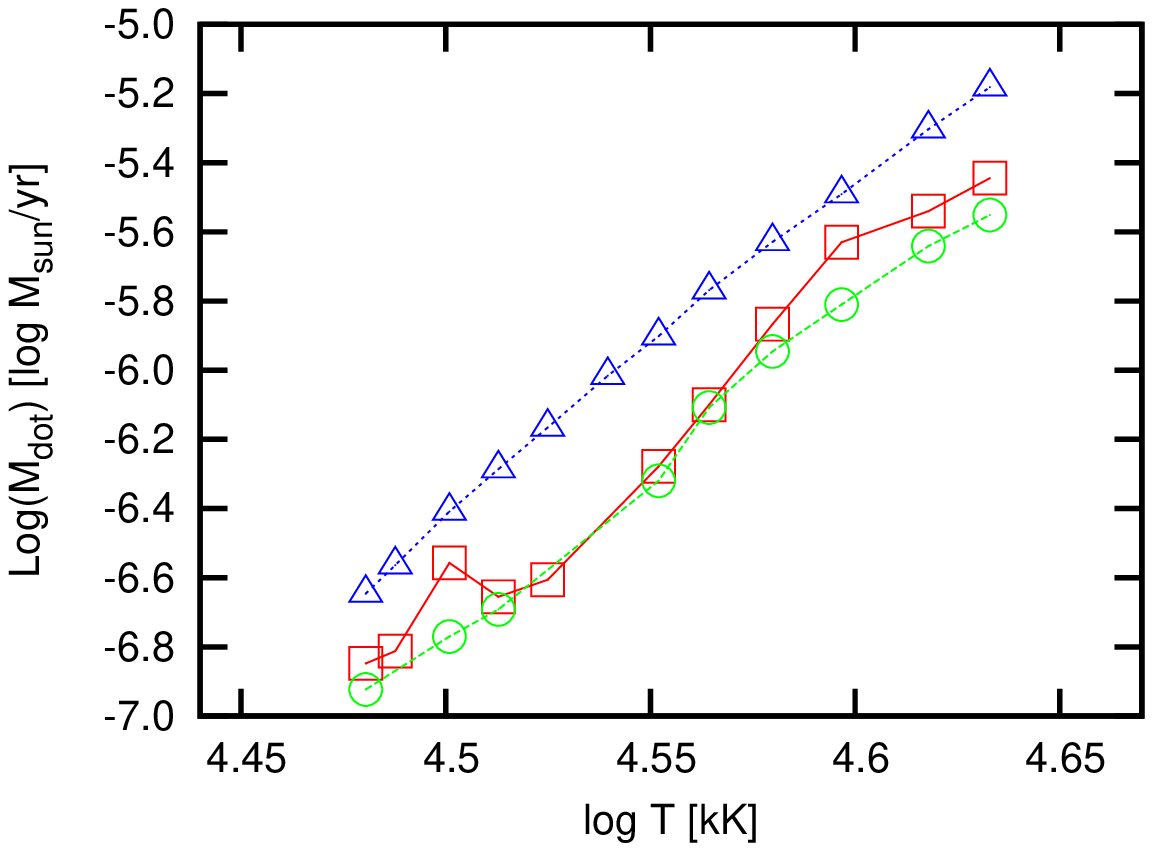}
     \includegraphics[width=0.60\textwidth,angle=0]{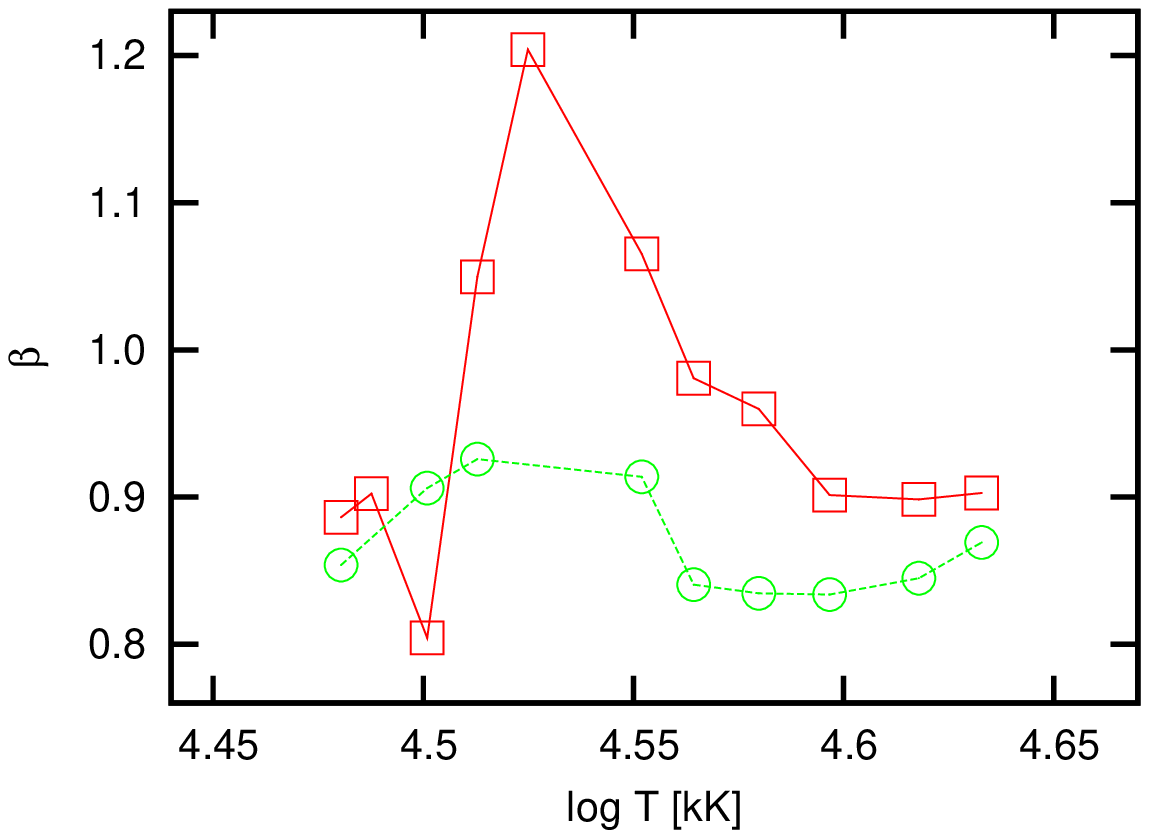} }
     
     \resizebox{10.3cm}{!}{
     \includegraphics[width=0.60\textwidth,angle=0]{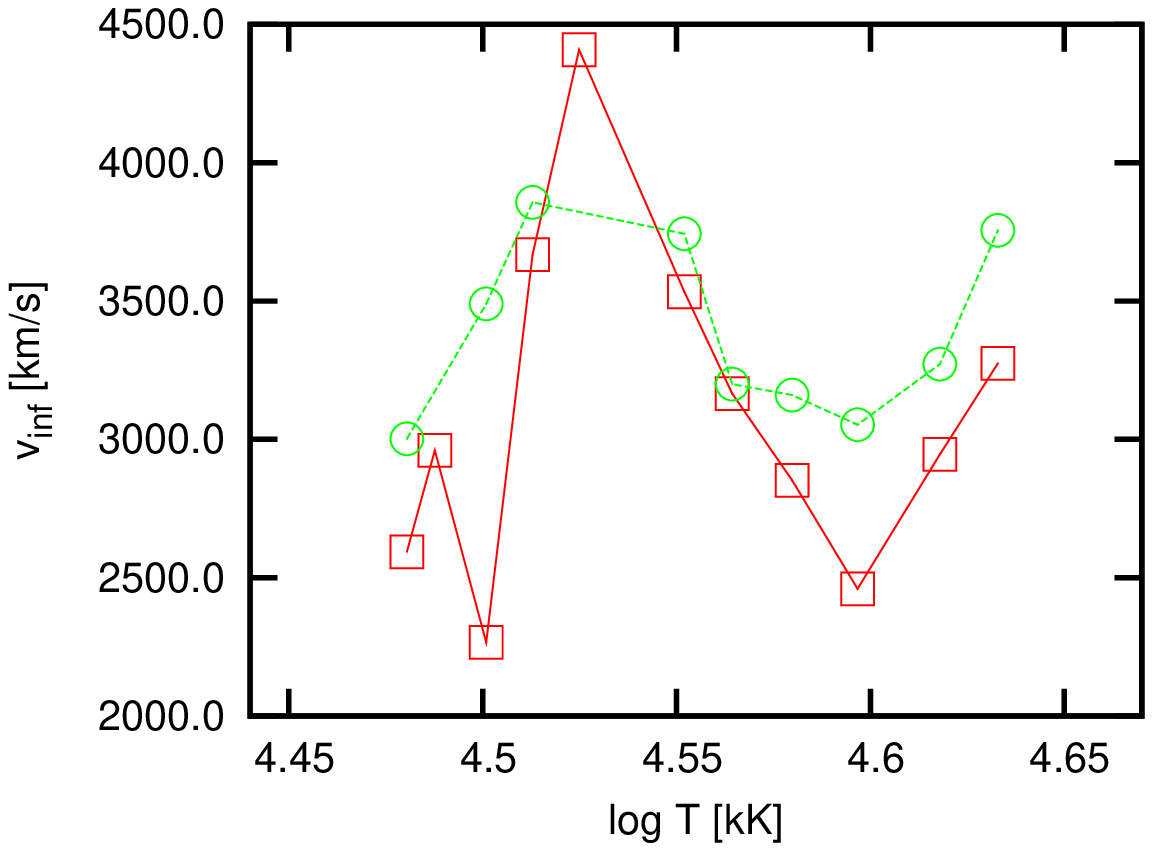}  
     \includegraphics[width=0.60\textwidth,angle=0]{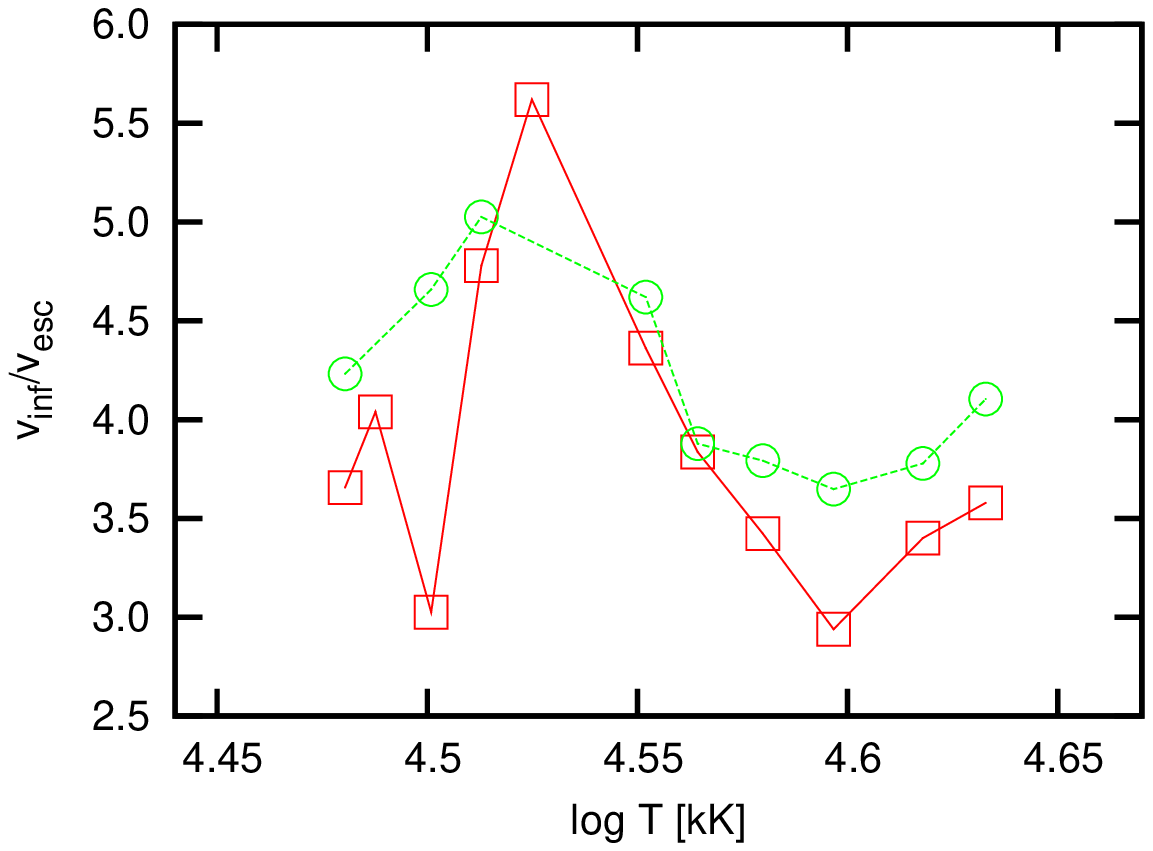} }    
  \caption{Predicted \mdot, \vinf ~and~ $\beta$ for giants. Colors have the same meaning as in Fig.~\ref{fig:masslossLV}.}
  \label{fig:masslossLIII}
\end{figure*}

\begin{figure*}
  \centering 
     \resizebox{10.3cm}{!}{
     \includegraphics[width=0.60\textwidth,angle=0]{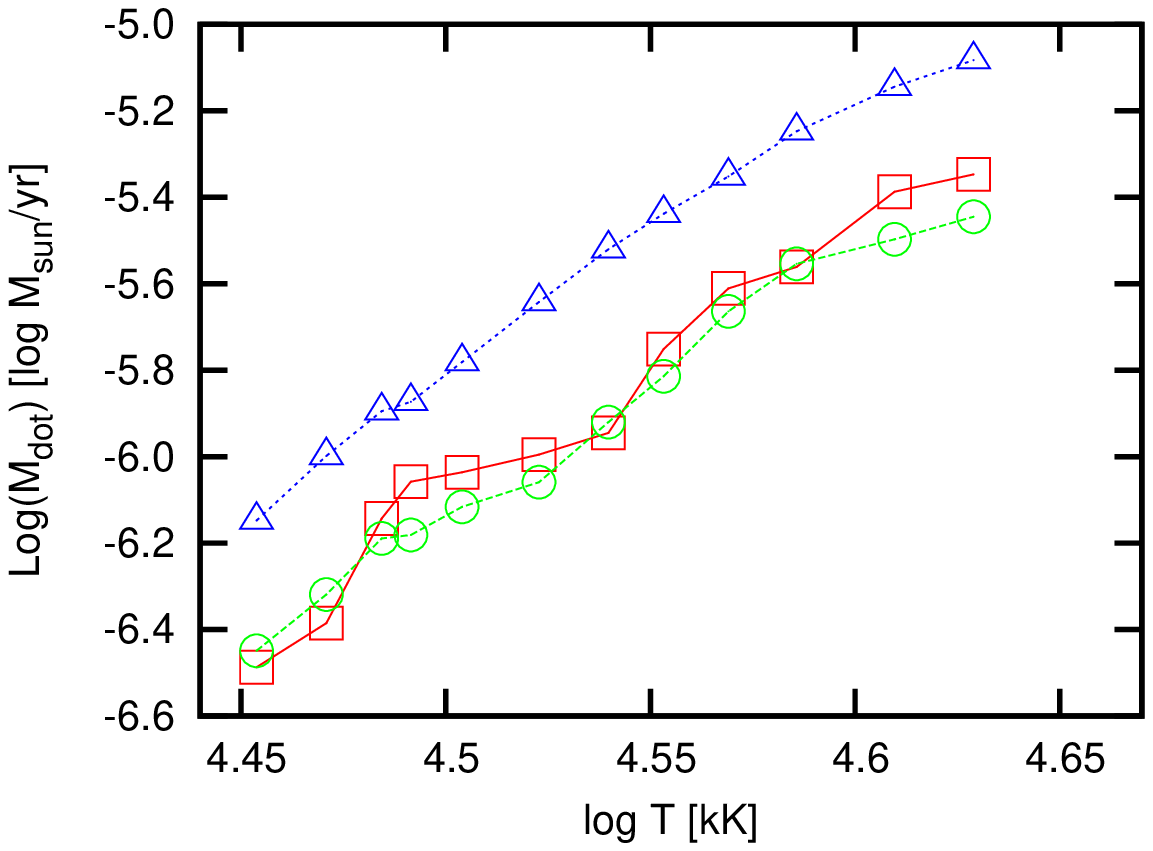}
     \includegraphics[width=0.60\textwidth,angle=0]{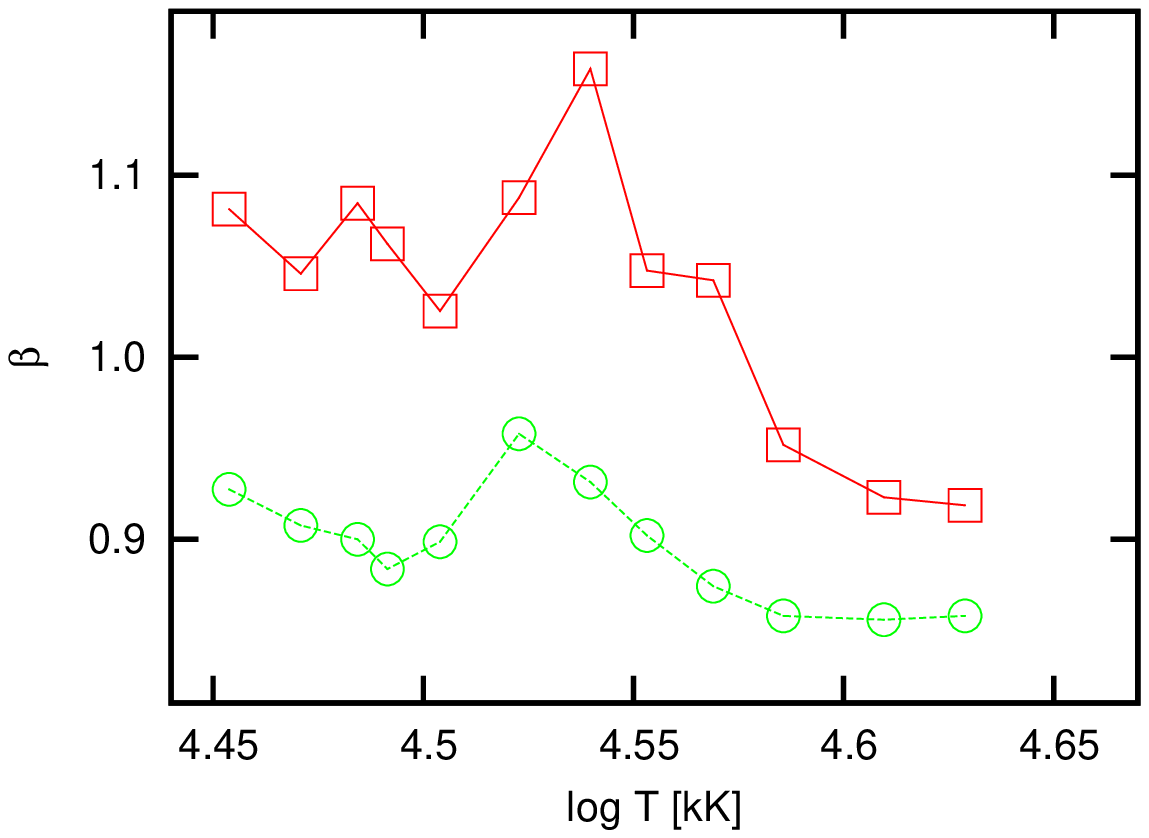} }
     
     \resizebox{10.3cm}{!}{
     \includegraphics[width=0.60\textwidth,angle=0]{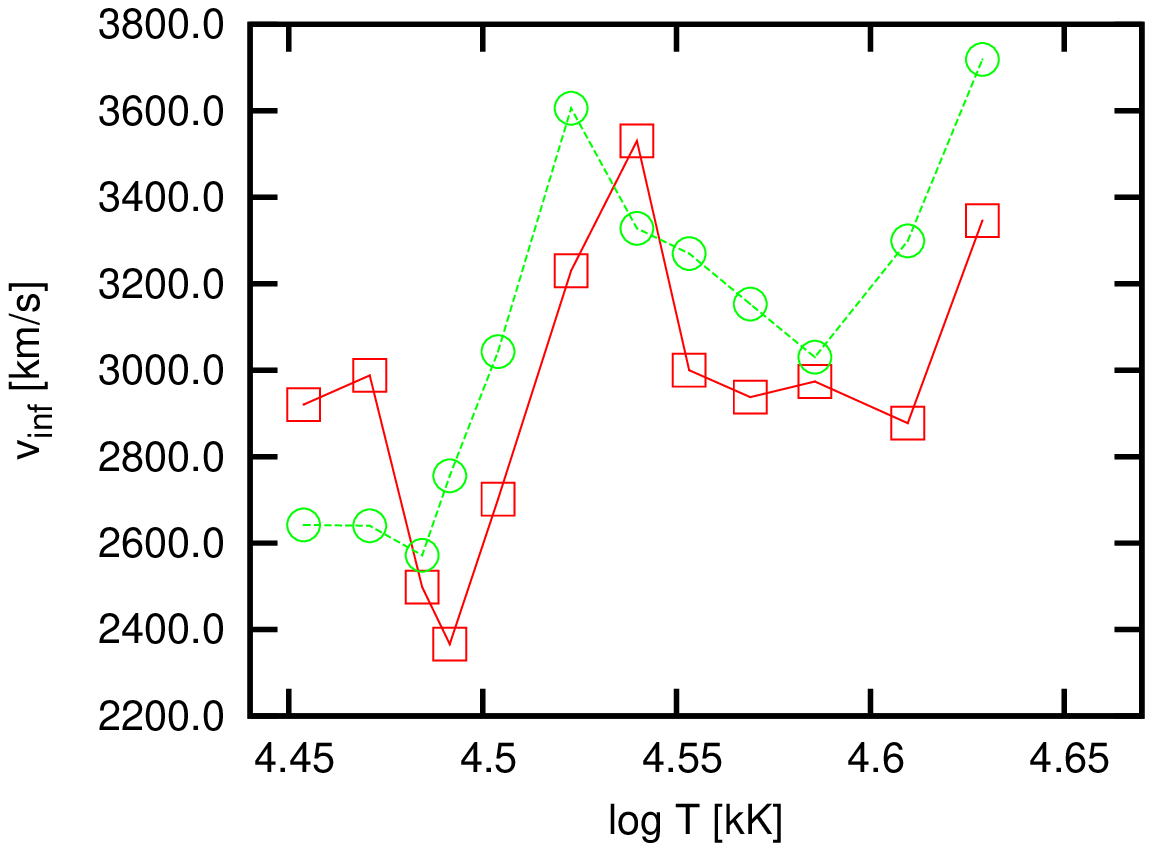}  
     \includegraphics[width=0.60\textwidth,angle=0]{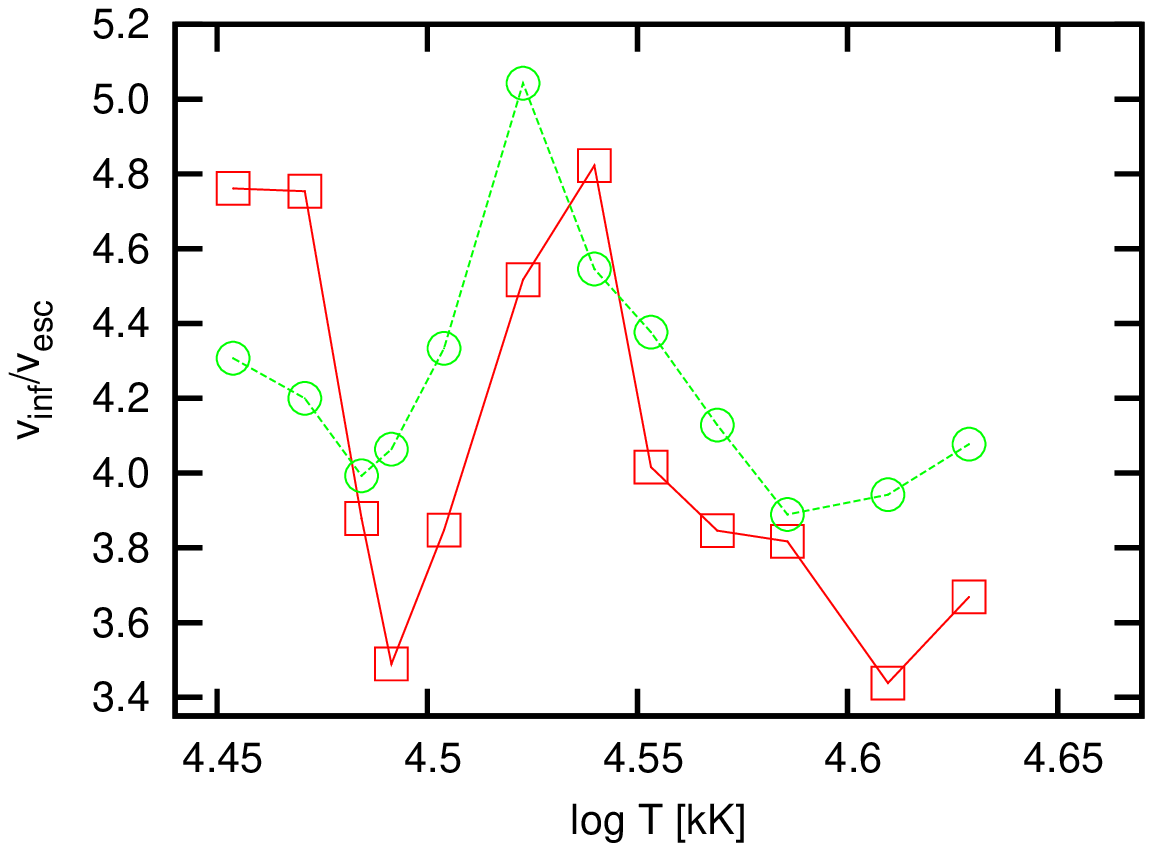} }    
  \caption{Predicted \mdot, \vinf ~and~ $\beta$ for supergiants. Colors have the same meaning as in Fig.~\ref{fig:masslossLV}.}
  \label{fig:masslossLI}
\end{figure*}

Figures \ref{fig:masslossLV}, \ref{fig:masslossLIII} 
and \ref{fig:masslossLI} show wind properties as a function of effective temperature, for dwarfs, giants and supergiants respectively.

We present all the results of the best beta method, i.e. prior to applying the requirement defined 
in Sect.~\ref{sec:sonicpoint1} that at the sonic point the acceleration due to the line force should 
approach the effective gravity. Having pointed this out, we first draw attention to the striking behavior
in our best-$\beta$ predictions of dwarfs. In the direction of decreasing temperature, the terminal
flow velocity drastically increases for spectral types O6.5 or later. For giants the O7 star shows a 
similar behavior. We argue below that this behavior reflects the failure of the wind to become
supersonic, therefore we interpret these solutions to be non-physical. 

\subsection{Early O-stars (spectral types O3 through O6)}

Let us, however, first focus on stars of spectral type earlier than O6.5. The two methods give
quite comparable results. The best-$\beta$ method predicts \mdot\ values 
that are higher by up to $\sim$0.1 to 0.3\,dex in all cases, i.e. dwarfs, giants and supergiants. 
The best-$\beta$ terminal flow velocities are $\sim$10 to 20 percent lower compared to the
hydrodynamic solutions. These
differences can be understood by focusing on the velocity structures near the sonic point.
In the best-$\beta$ solution the velocity law is steeper in the region near the sonic point,
therefore the sonic point is closer to the photosphere. This leads to a higher mass-loss 
rate and lower terminal velocity. 
The absolute value of the terminal velocity and the ratio of \vinf\ to the effective
escape velocity as a function of temperature will be compared to observations in 
Sect.~\ref{sec:discussion}. Typical error bars on the \vinf\ determination are
10 percent for the best-$\beta$ solutions (see Sect.~\ref{sec:bestbeta}) and 20 percent for the
exact solutions (see Sect.~\ref{sec:ns})
due to Monte Carlo noise on the line force (see also Fig.~\ref{fig:mcline}).

The slope of the velocity law in the best-$\beta$ solution increases slightly with
luminosity class, from typically 0.85 in dwarfs, to 0.95 in giants, to 1.0 in
supergiants. In the hydrodynamic models (method B) the $\beta_{\gamma}$ value is typically 
0.05-0.10 lower than the corresponding best-$\beta$ solution.

\begin{table*}[t]
\begin{center}
\caption{Model parameters, following \cite{2005A&A...436.1049M}, and predicted wind properties for dwarfs, giants and supergiants. 
         The label "spec" indicates that spectroscopic masses are adopted from \citeauthor{2005A&A...436.1049M}
         Predictions give the mass-loss rate, terminal
         velocities and $\beta$ parameters for both method A (best-$\beta$ solution) and B (hydrodynamic solution). 
         The 11th column states whether or not the best-$\beta$ solution fulfills the requirement that at the sonic point
         $\gline/\mathit{g}_{\rm eff} \simeq 1  $ (see Sect.~\ref{sec:sonicpoint1}). For the hydrodynamic solutions a failure of this
         requirement implies that we do not find a solution at all.
         The last column
         provides the $\mdot$ by~\cite{2001A&A...369..574V} when we use $\vinf = 2.6 \,v_{\rm esc}$ in their mass-loss recipe. 
         The hydrodynamic solution does not provide a $\beta$ value, but rather the fit parameter $\gamma$. 
         As to facilitate a comparison, we applied Eq.~\ref{eq:beta}
         to convert this $\gamma$ into a $\beta$.
         \label{table:grid-results}}
\tiny
\begin{tabular}{rrrrrrrrrrrrrrr}
\hline\\[-9pt] \hline\\[-7pt]
\multicolumn{7}{l}{Model Parameters} & \multicolumn{3}{l}{Method A} & \multicolumn{4}{l}{Method B} & Vink et al.\\[1pt]
ST & \teff\ & $\log \mathit{g}_{\rm spec}$  & $\log L$  & R & $M_{\rm spec}$  & $v_{\rm esc}$  & $\log \mdot$ & \vinf & $\beta$ & $\Gamma \simeq 1$ & $\log \mdot$ &  \vinf & $\beta_{\gamma}$ & $\log \mdot$ \\[1pt]
   & K      & cm s$^{-2}$       & \lsun   & \rsun &  \msun  & km/sec & log \msun/yr & km/sec & & at $v_{s}$ & \msun/yr & km/sec & & \msun/yr\\[1pt]
\hline\\[-7pt]
\multicolumn{14}{l}{{\em Dwarfs}} \\[1.5pt]
3   & 44616  & 3.92 & 5.83 & 13.84 &  58.34  & 1054  &  -5.641  &   3794  &  0.92  & yes & -5.972  &  4530  &  0.87  & -5.375 \\[1.5pt]
4   & 43419  & 3.92 & 5.68 & 12.31 &  46.16  & 1016  &  -5.836  &   3599  &  0.90  & yes & -5.929  &  3973  &  0.83  & -5.571 \\[1.5pt]
5   & 41540  & 3.92 & 5.51 & 11.08 &  37.28  &  992  &  -5.969  &   2838  &  0.84  & yes & -6.118  &  3394  &  0.79  & -5.829 \\[1.5pt]
5.5 & 40062  & 3.92 & 5.41 & 10.61 &  34.17  &  990  &  -6.152  &   2762  &  0.84  & yes & -6.265  &  3260  &  0.77  & -6.011 \\[1.5pt]
6   & 38151  & 3.92 & 5.30 & 10.23 &  31.73  &  994  &  -6.386  &   2697  &  0.81  & yes & -6.493  &  3277  &  0.77  & -6.234 \\[1.5pt]
6.5 & 36826  & 3.92 & 5.20 & 9.79  &  29.02  &  983  & [-7.243] &  [6395] & [1.87] &  no & -6.918  &  5244  &  0.95  & -6.427 \\[1.5pt]
7   & 35531  & 3.92 & 5.10 & 9.37  &  26.52  &  972  & [-7.340] &  [7325] & [2.65] &  no &      -- &    --  &    --  & -6.624 \\[1.5pt]
7.5 & 34419  & 3.92 & 5.00 & 8.94  &  24.15  &  959  & [-7.745] & [12028] & [1.81] &  no &      -- &    --  &    --  & -6.820 \\[1.5pt]
8   & 33383  & 3.92 & 4.90 & 8.52  &  21.95  &  944  & [-7.781] & [10650] & [1.57] &  no &      -- &    --  &    --  & -7.019 \\[1.5pt]
8.5 & 32522  & 3.92 & 4.82 & 8.11  &  19.82  &  923  & [-7.802] &  [9427] & [1.40] &  no &      -- &    --  &    --  & -7.167 \\[1.5pt]
9   & 31524  & 3.92 & 4.72 & 7.73  &  18.03  &  908  & [-7.818] &  [8283] & [1.21] &  no &      -- &    --  &    --  & -7.374 \\[1.5pt]
9.5 & 30488  & 3.92 & 4.62 & 7.39  &  16.46  &  892  & [-7.793] &  [6704] & [1.10] &  no &      -- &    --  &    --  & -7.590 \\[1.5pt]
\hline\\[-7pt]
\multicolumn{14}{l}{{\em Giants}} \\[1.5pt]
3   & 42942  & 3.77 & 5.92 & 16.57 &  58.62  &  915  &  -5.445  &   3275  &  0.90  & yes & -5.551  &  3756  &  0.87  & -5.182 \\[1.5pt]
4   & 41486  & 3.73 & 5.82 & 15.83 &  48.80  &  866  &  -5.540  &   2945  &  0.90  & yes & -5.641  &  3272  &  0.84  & -5.303 \\[1.5pt]
5   & 39507  & 3.69 & 5.70 & 15.26 &  41.48  &  837  &  -5.630  &   2460  &  0.90  & yes & -5.810  &  3053  &  0.83  & -5.491 \\[1.5pt]
5.5 & 38003  & 3.67 & 5.63 & 15.13 &  38.92  &  833  &  -5.867  &   2852  &  0.96  & yes & -5.946  &  3160  &  0.83  & -5.629 \\[1.5pt]
6   & 36673  & 3.65 & 5.56 & 14.97 &  36.38  &  825  &  -6.100  &   3165  &  0.98  & yes & -6.108  &  3200  &  0.84  & -5.769 \\[1.5pt]
6.5 & 35644  & 3.63 & 5.49 & 14.74 &  33.68  &  810  &  -6.278  &   3534  &  1.07  & yes & -6.320  &  3743  &  0.91  & -5.902 \\[1.5pt]
7   & 34638  & 3.61 & 5.43 & 14.51 &  31.17  &  798  & [-6.804] &  [7140] & [3.46] &  no &    --   &    --  &    --  & -6.016 \\[1.5pt]
7.5 & 33487  & 3.59 & 5.36 & 14.34 &  29.06  &  785  &  -6.606  &   4408  &  1.20  & yes &    --   &    --  &    --  & -6.166 \\[1.5pt]
8   & 32573  & 3.57 & 5.30 & 14.11 &  26.89  &  768  &  -6.655  &   3668  &  1.05  & yes & -6.692  &  3857  &  0.93  & -6.286 \\[1.5pt]
8.5 & 31689  & 3.55 & 5.24 & 13.88 &  24.84  &  749  &  -6.557  &   2266  &  0.80  & yes & -6.770  &  3490  &  0.91  & -6.409 \\[1.5pt]
9   & 30737  & 3.53 & 5.17 & 13.69 &  23.07  &  733  &  -6.812  &   2960  &  0.90  & yes &    --   &    --  &    --  & -6.564 \\[1.5pt]
9.5 & 30231  & 3.51 & 5.12 & 13.37 &  21.04  &  709  &  -6.848  &   2594  &  0.89  & yes & -6.923  &  3002  &  0.85  & -6.646 \\[1.5pt]
\hline\\[-7pt]
\multicolumn{14}{l}{{\em Supergiants}} \\[1.5pt]
3   & 42551  & 3.73 & 6.00 & 18.47 &  66.89  &  912  & -5.347 &  3346  & 0.92 & yes & -5.445 & 3719  & 0.86  & -5.083 \\[1.5pt]
4   & 40702  & 3.65 & 5.94 & 18.91 &  58.03  &  837  & -5.387 &  2877  & 0.92 & yes & -5.497 & 3299  & 0.86  & -5.144 \\[1.5pt]
5   & 38520  & 3.57 & 5.87 & 19.48 &  50.87  &  779  & -5.561 &  2974  & 0.95 & yes & -5.554 & 3030  & 0.86  & -5.247 \\[1.5pt]
5.5 & 37070  & 3.52 & 5.82 & 19.92 &  48.29  &  764  & -5.611 &  2938  & 1.04 & yes & -5.664 & 3153  & 0.87  & -5.352 \\[1.5pt]
6   & 35747  & 3.48 & 5.78 & 20.33 &  45.78  &  747  & -5.751 &  3000  & 1.05 & yes & -5.814 & 3270  & 0.90  & -5.438 \\[1.5pt]
6.5 & 34654  & 3.44 & 5.74 & 20.68 &  43.10  &  732  & -5.945 &  3531  & 1.16 & yes & -5.920 & 3328  & 0.93  & -5.520 \\[1.5pt]
7   & 33326  & 3.40 & 5.69 & 21.14 &  40.91  &  715  & -5.995 &  3230  & 1.09 & yes & -6.059 & 3606  & 0.96  & -5.642 \\[1.5pt]
7.5 & 31913  & 3.36 & 5.64 & 21.69 &  39.17  &  702  & -6.036 &  2702  & 1.03 & yes & -6.116 & 3043  & 0.90  & -5.781 \\[1.5pt]
8   & 31009  & 3.32 & 5.60 & 22.03 &  36.77  &  678  & -6.058 &  2366  & 1.06 & yes & -6.181 & 2756  & 0.88  & -5.873 \\[1.5pt]
8.5 & 30504  & 3.28 & 5.58 & 22.20 &  33.90  &  644  & -6.143 &  2498  & 1.08 & yes & -6.189 & 2572  & 0.90  & -5.895 \\[1.5pt]
9   & 29569  & 3.23 & 5.54 & 22.60 &  31.95  &  629  & -6.385 &  2988  & 1.05 & yes & -6.319 & 2640  & 0.91  & -5.998 \\[1.5pt]
9.5 & 28430  & 3.19 & 5.49 & 23.11 &  30.41  &  613  & -6.487 &  2921  & 1.08 & yes & -6.449 & 2642  & 0.93  & -6.148 \\[1.5pt]
\hline\\[-7pt]
\end{tabular}
\end{center}
\normalsize
\end{table*}

\subsection{Late O-stars (spectral types O6.5 through O9.5)}
\label{sec:lateO}

\begin{figure}
     \resizebox{8cm}{!}{
     \includegraphics[width=0.70\textwidth,angle=0]{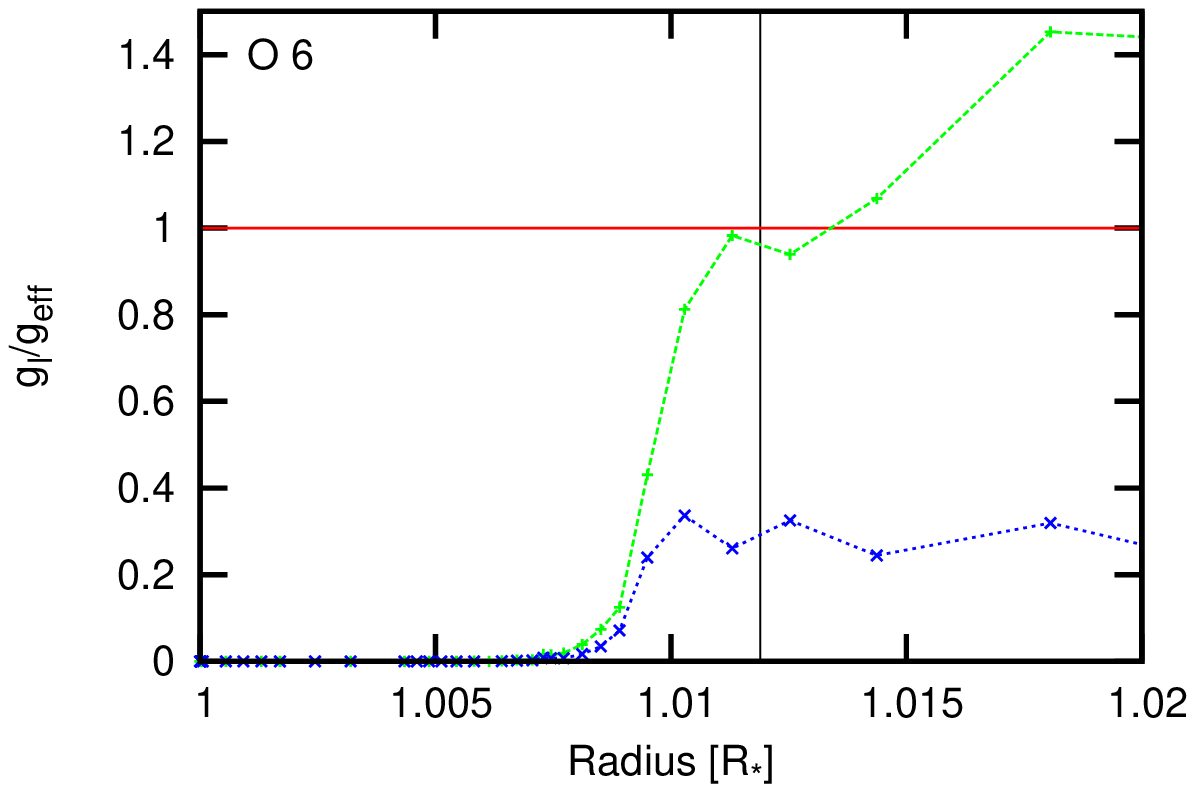}  }
     \resizebox{8cm}{!}{
     \includegraphics[width=0.70\textwidth,angle=0]{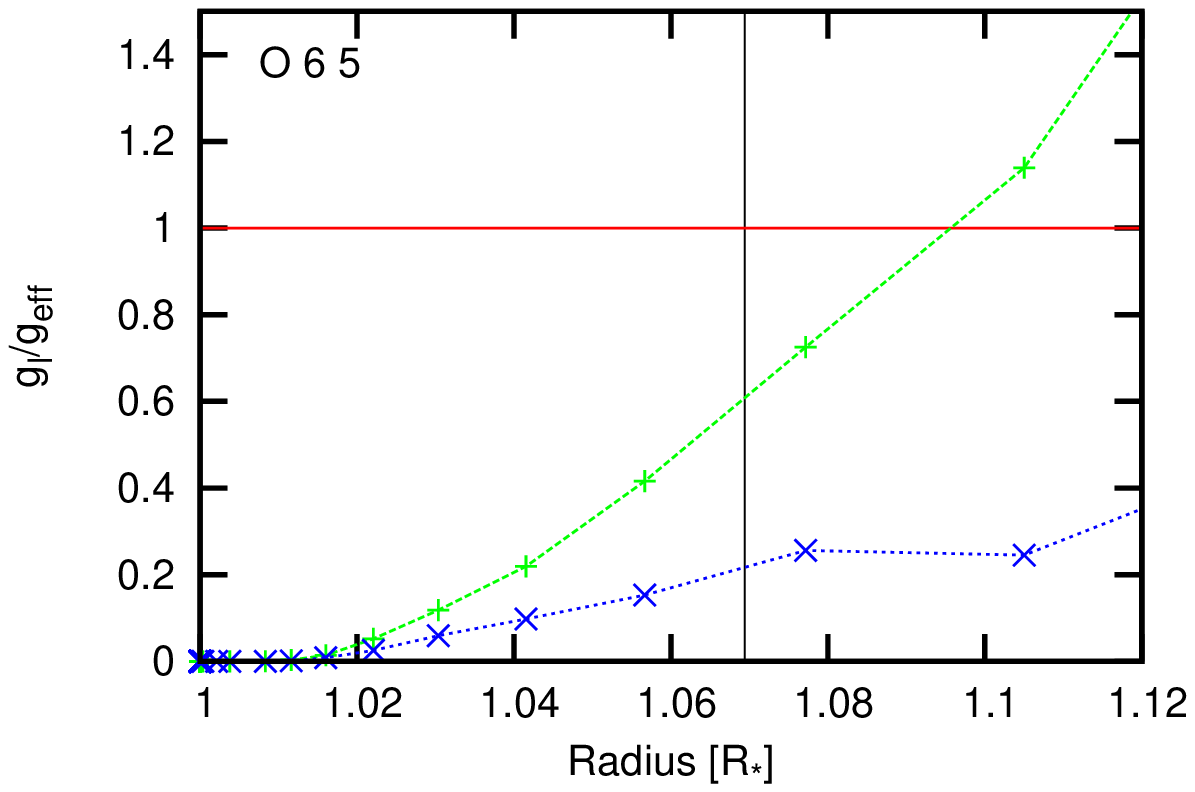} }
      \caption{The Monte Carlo line force as a fraction of the effective gravity in the
               region around the sonic point for the best-$\beta$ solution of the O6\,V (top
               panel) and O6.5\,V (bottom panel) stars. The contribution of iron is shown 
               separately (blue dotted line). Note that in the case of the O6.5 star the 
               line acceleration does not balance the effective gravity at the sonic point. 
               This is interpreted as a failure to support a line driven wind.}
  \label{fig:sonicpoint}
\end{figure}

Figures~\ref{fig:masslossLV} and~\ref{fig:masslossLIII} show that for spectral type O6.5 the terminal
velocity of dwarfs and giants suddenly peaks, relative to spectral type O6. We investigate this
behavior in more detail in Fig.~\ref{fig:sonicpoint} in which the line force from the Monte Carlo
simulation is plotted in the region around the sonic point for the best-$\beta$ solutions of the
dwarf O6 and O6.5 star. For the O6 star, the line force at the base of the wind (below the sonic
point) rises steeply. {\em At first the dominant contributors are iron lines, notably from 
Fe\,{\sc v}}. The ensemble of transitions mainly occur between excited states, although some are from 
meta-stable states that are relatively strongly populated. Further out, the iron contribution 
levels out (at $\sim 30$\%) and other elements start to contribute to the force, such as carbon, nitrogen,
sulfur, argon and nickel. The contribution of resonance lines of carbon and nitrogen at the sonic 
point amounts to $\sim$ 20\%. Note that at the sonic point the $\gline/\mathit{g}_{\rm eff} \sim 1$
condition is nicely fulfilled for the O6\,V.

For the O6.5 star this is not the case. Here the line force at the base of the wind (below the sonic
point) rises only gradually. The difference with the O6\,V star is that in this region iron is 
mainly in the form of Fe\,{\sc iv}, which for this particular spectral flux distribution 
is less efficient in absorbing stellar flux than are
Fe\,{\sc v} lines\footnote{We note that a similar situation occurs at spectral type B1, where 
the relatively inefficient Fe {\sc iv} lines are replaced by the more effective Fe {\sc iii} lines 
\citep{1999A&A...350..181V}.}. Therefore the velocity structure will be shallower, limiting the potential
of other elements in contributing to the force. As a result the sonic point starts to shift out
to larger radii, and we find that at the sonic point the cumulative line acceleration is some 40\%
less than the effective gravity. We therefore interpret this outcome as a failure of the wind
to become supersonic at $r_{\rm s}$ and do not consider it to be a physical solution.

The best-$\beta$ solutions where we clearly encounter this problem have brackets placed around the
predicted wind properties as listed in Table~\ref{table:grid-results}. 
These include all the dwarf stars of spectral type O6.5 or later.
They are to be considered non-physical. 

The supergiants do not suffer from this problem. In all cases
the $\gline/\mathit{g}_{\rm eff} \sim$\,1 was reached
at the sonic point and we consider them physical solutions. The terminal velocities for the 
O6.5\,I to O9.5\,I scatter by about 20\%, with a small hint that here also the O6.5 star has a 
higher \vinf. The latter occurs because elements such as silicon, iron and sulfur add
to the line force in the outer wind along with the normal contribution of carbon, nitrogen and oxygen. 

The value of $\beta$ for the late spectral types increases
to 1.05 from 1.0 for earlier spectral types. The $\beta_{\gamma}$ values associated to the 
hydrodynamic solutions increase marginally compared to that in early-O stars. 

\section{Discussion}
\label{sec:discussion}
In discussing our results we first compare with previous theoretical predictions for mass-loss rates 
and terminal velocities in sections \ref{sec:jorickmassloss} and \ref{sec:MCAK-theory}. We compare to 
observations in section \ref{sec:Observations}.

\subsection{Comparison to Vink et al. mass-loss recipe}
\label{sec:jorickmassloss}

The Monte Carlo method by \citet{1997ApJ...477..792D} as summarized in Sect.~\ref{sec:Method} 
has been used by \citet{2000A&A...362..295V,2001A&A...369..574V} to compute a grid of 
mass-loss rates for O-type stars from which a fitting formula has been derived that 
provides \mdot\ as a function of luminosity, effective temperature, mass and the
ratio of the terminal velocity over the effective escape velocity, i.e. 
$\vinf/\vesc$. This mass loss prescription is widely used in stellar evolution 
predictions \citep[see e.g][]{2003A&A...404..975M,2005A&A...443..243P,
2006ApJ...647..483L,2006A&A...452..295E,2009CoAst.158...55B,2010A&A...512L...7V}. 

\begin{figure*}[t!]
     \centering
       \resizebox{18cm}{!}{
     \includegraphics[width=0.70\textwidth,angle=0]{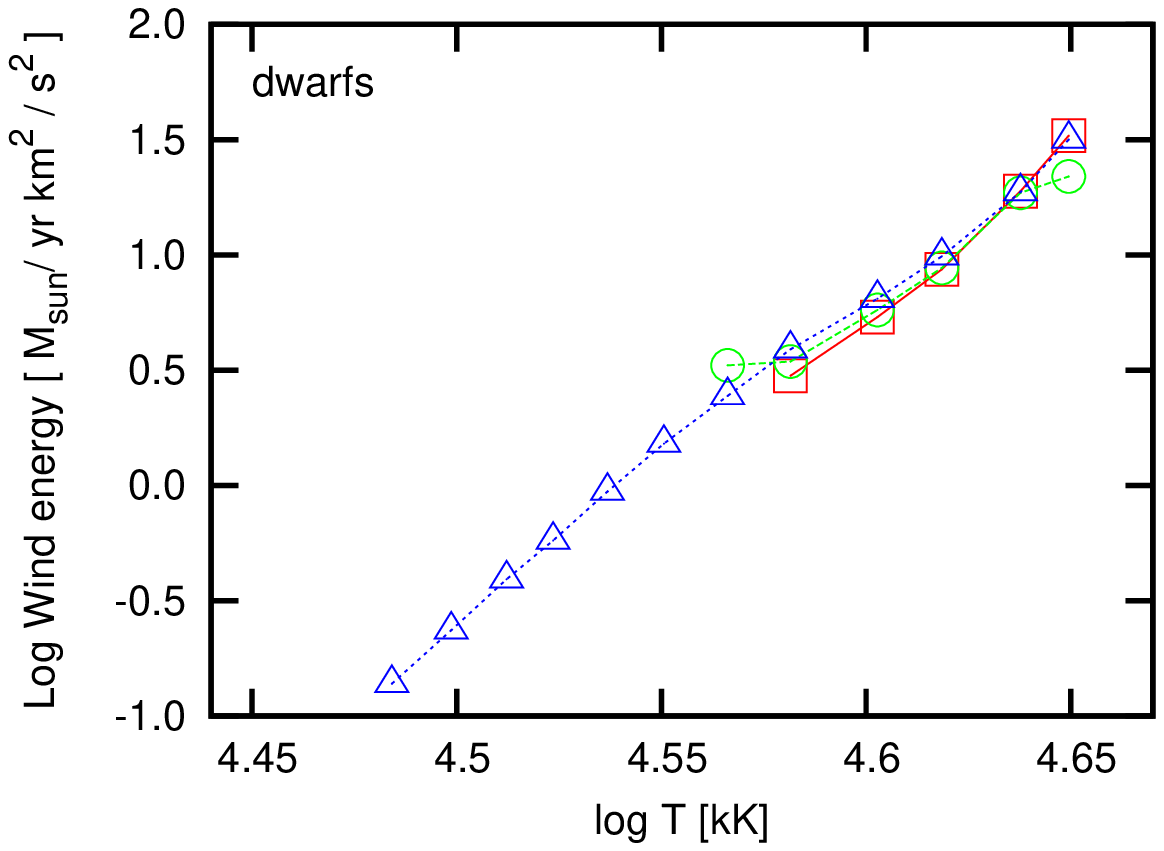}       
     \includegraphics[width=0.70\textwidth,angle=0]{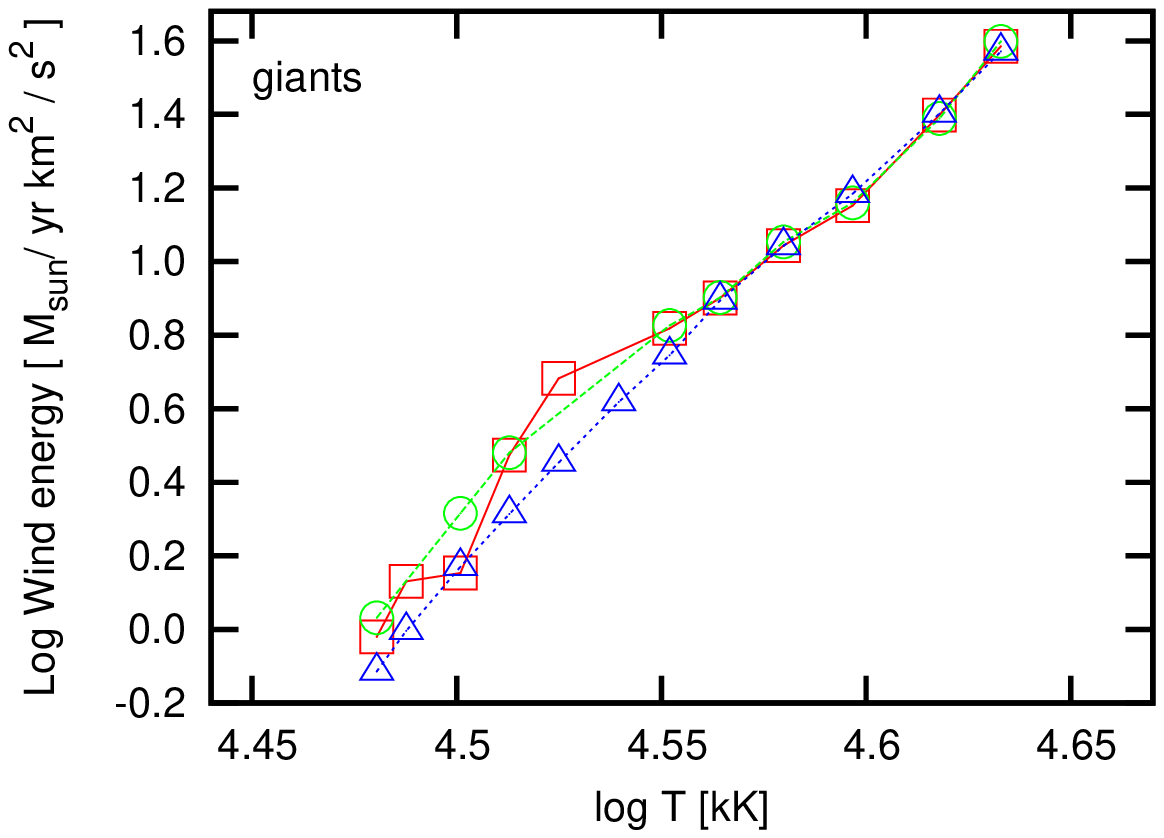}
     \includegraphics[width=0.70\textwidth,angle=0]{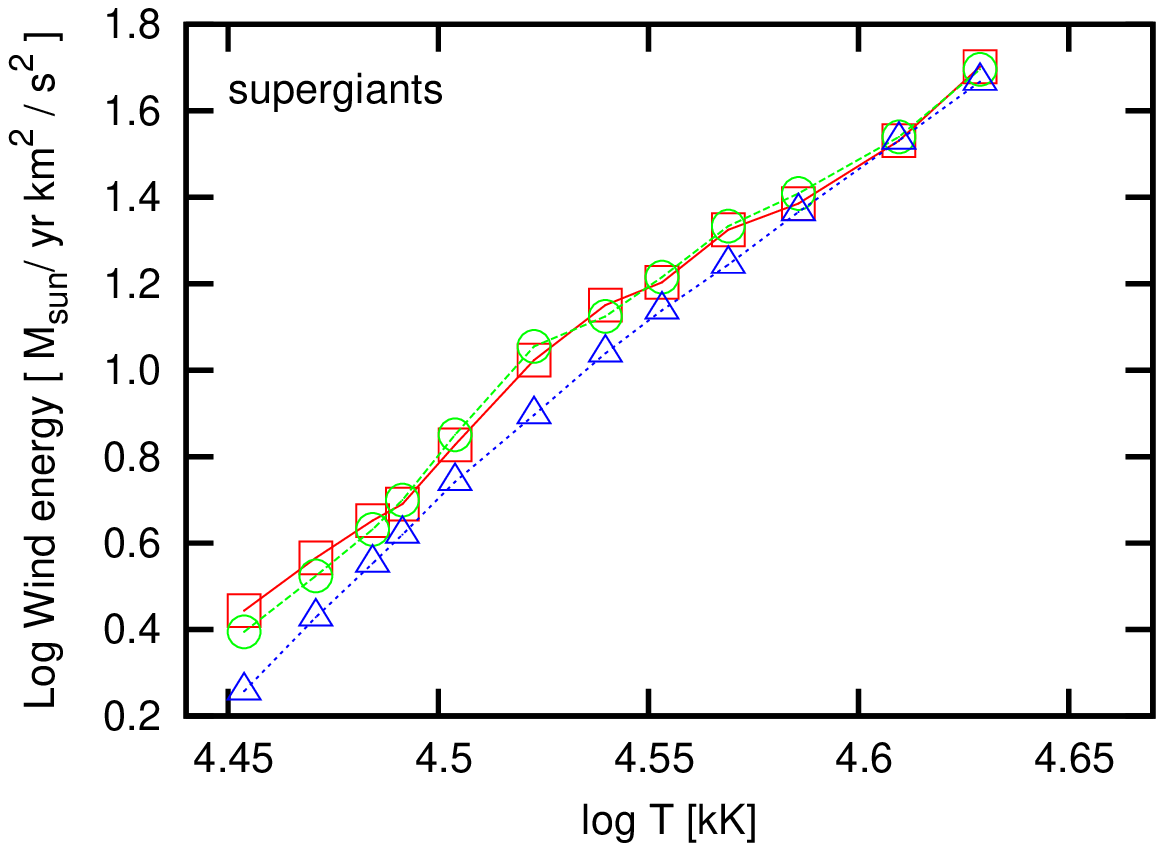}     
      }
  \caption{The wind energy as a function of the effective temperature for dwarfs 
          (left panel), giants (middle panel) and supergiants (right panel).  The best-$\beta$ method is 
	  given by the red squares, the hydrodynamic method by the green circles and the \citeauthor{2000A&A...362..295V}-recipe
	  by the blue triangles. Note that the kinetic energy in the wind is almost equal for all three 
          methods. The mass-loss rate and terminal velocity for the three methods vary.
          \label{fig:windenergy}}
\end{figure*}

For the spectral range that is investigated here the canonical value, based on empirical findings, for the ratio
of terminal velocity over escape velocity is 2.6. To compare to the \cite{2000A&A...362..295V}
results, we calculated the mass-loss rate of our grid of stars using  
their prescription, that assumes $\beta = 1$. The results 
are given in the last column of Table~\ref{table:grid-results}. Figure~\ref{fig:windenergy} 
shows the total energy that is extracted from the radiation field and that is transferred
to the stellar wind for all three methods: best-$\beta$, hydrodynamic and 
\citeauthor{2000A&A...362..295V} prescription. All three methods yield similar,
but not identical results in the regime where the best-$\beta$ and hydrodynamical method 
provide physical solutions. In terms of mass loss rates, we find that the predictions 
with the best-$\beta$ and hydrodynamical method are on average about 0.2 to 0.3 dex lower 
than \citeauthor{2000A&A...362..295V}, again with the clear exception of the stars for 
which we fail to drive a stellar wind. As suggested by the similar wind energies
the terminal velocities predicted by our best-$\beta$ and hydrodynamical method turn out
to be higher than adopted by \citeauthor{2000A&A...362..295V}. We discuss these
\vinf\ in more detail in Sect.~\ref{sec:Observations} as well as the reason why
\citeauthor{2000A&A...362..295V} are able to predict \mdot\ values for late O-type
dwarfs and giants, where we fail.

\emph{We emphasize that if the \citeauthor{2000A&A...362..295V} prescription is
used assuming the terminal velocities predicted by our best-$\beta$ or hydrodynamical
method, wherever these yield physical solutions, the mass loss rates agree
to within $\sim$0.1 dex.}

\subsection{Comparison to (Modified)CAK-theory}
\label{sec:MCAK-theory}

Since \citet{1970ApJ...159..879L} it is generally accepted that the winds of massive 
stars are driven by the transfer of momentum (and energy) from the radiation field to 
the atmospheric gas, and that atomic transitions play a pivotal role in this process. 
\cite{1975ApJ...195..157C} 
describe the force associated to atomic transitions by introducing a force multiplier 
\beq
\label{eq:cak}
M(t) = \frac{\gline(t)}{\mathit{g}_{\rm e}(t)} = k\,t^{-\alpha},
\eeq 
where $k$ and $\alpha$ are fitting parameters and $t$ is an optical depth like
parameter given by:
\beq
\label{eq:tforce}
t = \sigma_{\rm e} \,\rho\, v_{\rm th} \left( \frac{dv}{dr} \right)^{-1}. 
\eeq 
Here $\rho$ is the density, $v_{\rm th}$ the thermal velocity of carbon ions at 
the effective temperature of the star \citep{1986A&A...164...86P}  and $\sigma_{\rm e}$
the mass scattering coefficient of the free electrons.
This parametrization of 
the line force is based on the expression of the force multiplier for a single spectral line,
\beq
\label{eq:cakline}
M_{\rm line}(t) = \frac{\Delta\nu_{\rm D} \,F_{\nu}}{F} \frac{1}{t} \left[ 1-\exp{(\eta t)} \right],
\eeq 
where $\Delta \nu_{\rm D}$ is the Doppler shift of the frequency of the spectral line due 
to the thermal velocity of the particles in the wind, $F_{\nu}$ is the flux at 
frequency $\nu$, $F$ the total flux and $\eta$ is the ratio of line opacity to electron scattering 
opacity. 
Note that for optically thin lines $M_{\rm line}(t)$ becomes independent of
$t$, whilst for optically thick lines $M_{\rm line}(t) \propto t$. The cumulative
effect of an ensemble of lines of various strengths is then expressed by 
Eq.~\ref{eq:cak}.
The constant $\alpha$ in this expression is a measure of the ratio of line acceleration from optically thick lines only to 
the total line acceleration and $k$ is related to the overall (line)strength of the ensemble of lines. See \cite{2000A&AS..141...23P} for a more in depth discussion on  the CAK
line force.
It is assumed that $\alpha$ and $k$ are constants throughout the wind
\citep[but see][]{2002ApJ...577..389K}.
The effects of changes in the ionization structure of the wind are modeled by multiplying 
expression~\ref{eq:cak} by the term $(n_{\rm e}/W)^{\delta}$, 
introduced by \citet{1982ApJ...259..282A}. $n_{\rm e}$ is the electron number density, 
$W$ the dilution factor and $\delta$ is a constant.

The parametrization of the line force as given by Eq. \ref{eq:cak} leads to an analytical 
expression for $\mdot$ and $\vinf$ as a function of the fitting parameters $k$ and $\alpha$ 
and the stellar parameters. These expressions are given by \cite{1975ApJ...195..157C}.
\cite{1986A&A...164...86P} extend these expressions to account for the finite size of the 
stellar disk.

To allow for a comparison with Abbott's results for the behavior of the force multiplier
as a function of $t$, we ignore the effect of $n_{\rm e}/W$.
We calculated the force multiplier of the simulated line force, i.e. Eq.~\ref{eq:grad}, at all 
our radius grid points and determined the corresponding value $t$. 
Typically, the optical depth like parameter ranges from $t = 10^{-5}$ to large $t$. At large $t$ 
the line force can be neglected compared to the continuum radiation force.
Following \cite{1982ApJ...259..282A}, we do not consider these large $t$ points in this discussion
but we focus on the range $t < 10^{-0.5}$. 

\begin{figure}[b!]
    \begin{center}
    \resizebox{8cm}{!}{
     \includegraphics[width=0.70\textwidth,angle=0]{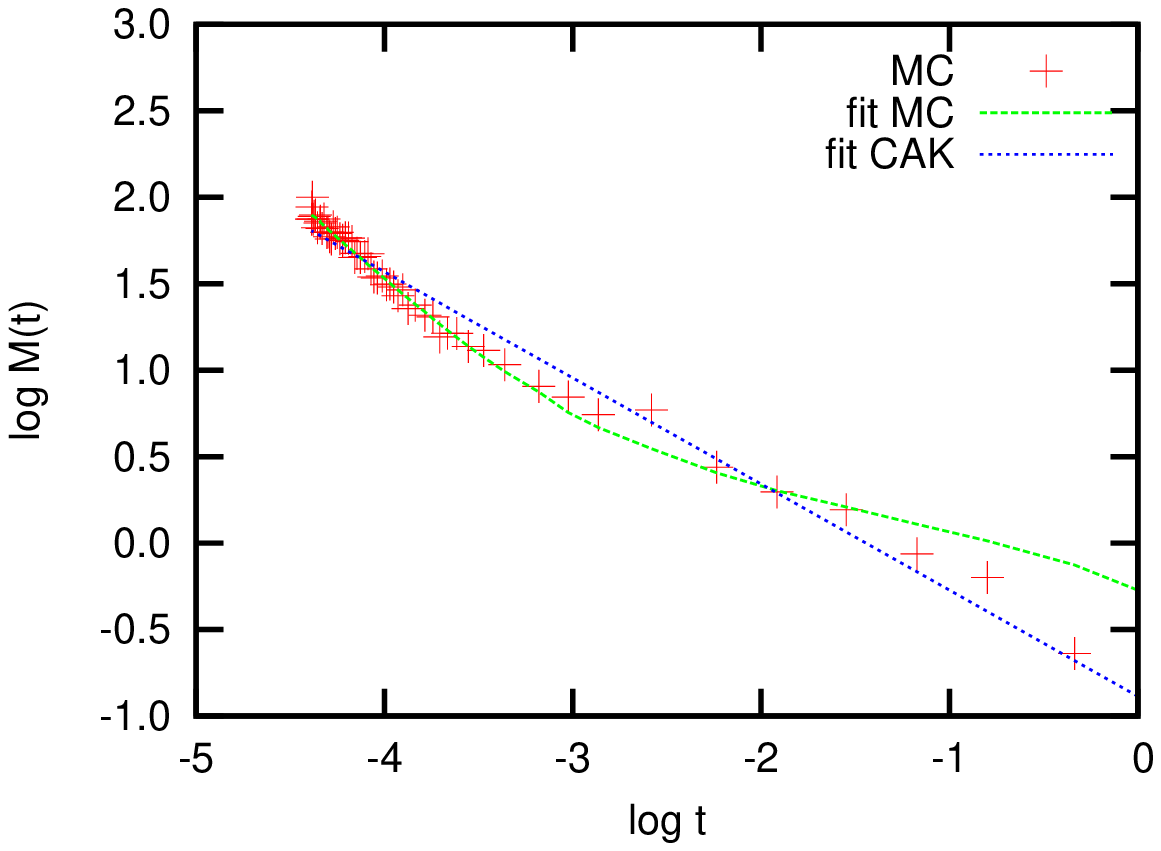}}
    \resizebox{8cm}{!}{
     \includegraphics[width=0.70\textwidth,angle=0]{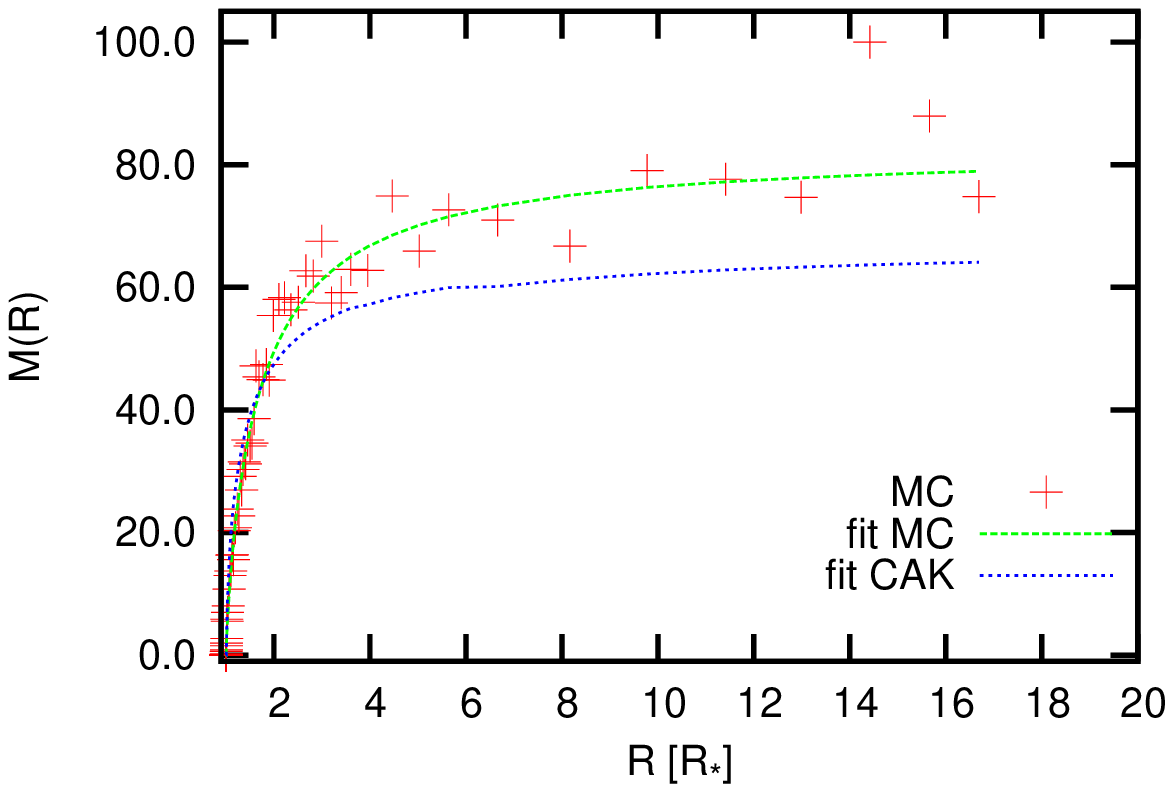}
     }
   \end{center}
   \caption{{\em Top panel:} 
            The logarithm of the force multiplier $M(t)$ for the hydrodynamic solution of our O3\,V model (red
            plusses) as a function of the optical depth like parameter $t$; our fit function Eq.~\ref{eq:grad} to this data (green line)
            and the CAK fit function Eq.~\ref{eq:cak} to this data (blue line). 
            {\em Bottom panel:} The force multiplier as a function of radius $r$ for
            the same star. Note that the CAK force multiplier is smaller than ours 
            for large radii, resulting in a lower predicted $\vinf$.}  
   \label{fig:cakforcemultiplyer}
\end{figure}

Figure~\ref{fig:cakforcemultiplyer} compares the behavior of the force multiplier for our 
best-$\beta$ solution of the O3\,V star. 
{\em Note that our Monte Carlo solution shows that $M(t)$ can not be fully described by a strict
power law, as assumed in (modified) CAK}, or equivalently, $\alpha$ is not independent of
$t$ \citep{2000PhDT........45V}. 
Two causes can be pointed out \citep[see e.g.][]{1987A&A...184..227P,1997A&A...322..598S}: {\em i)}
the presence of a diffuse field due to multiple scatterings; {\em ii)} a complex
behavior of radial stratification of the excitation and ionization, specifically
near the sonic point. Here the wind accelerates rapidly, which causes a sudden
steep drop in the electron density. As a result, elements which happen to have two 
dominant ionization stages (near $r_{\rm s}$) may temporarily re-ionize.

The introduction of a $\delta$-dependence of the CAK force multiplier in Eq.~\ref{eq:cak},
by adding the term $(n_{\rm e}/W)^{\delta}$, does not improve the fit to the Monte
Carlo line force. This extended CAK description describes a plane
through the three dimensional space spanned by the logarithms of $t$, $n_{\rm e}/W$ and $M(t)$, while
the Monte Carlo line force follows a curved line through this space and is thus not confined to that plane. 

Our fitting function, Eq.~\ref{eq:grad}, nicely captures the curved behavior of the line
force in the supersonic part of the flow.\footnote{We also compared our fitting formula to the output of the starting model of 
the iteration cycle of the O3\,V star. This allows us to assess whether 
the iteration procedure perhaps {\em forces} the line force into the shape
of Eq.~\ref{eq:grad}. However, also the line force resulting from the first
iteration cycle is well represented by our description, indicating that it
is quite generic.} 

Using the force multipliers $k$ and $\alpha$ as derived from the Monte Carlo
line force we can compute the CAK mass loss and terminal velocity \citep{1975ApJ...195..157C}. The 
$\mdot_{\rm CAK}$  values derived from these $k$ and $\alpha$ are typically 0.0 to 0.3 dex higher than our best-$\beta$ results and
our hydrodynamic solutions. 
A comparison of the terminal velocities is not meaningful since the slope in the supersonic part of the wind
is not represented well by $\alpha$. Therefore, the velocities derived with our $k$ and $\alpha$ are on the order of the escape velocity. If we compare to the modified CAK 
terminal velocities, following \citep{1986A&A...164...86P}, we note that they are slightly lower than the velocities we derive (see also section ~\ref{sec:Observationtv}).

\subsection{Comparison with observations}
\label{sec:Observations}
In this section, we compare our results to observations. We first compare predicted
and empirical terminal velocities. Given that we find that for stars more luminous than
$10^{5.2}$\,\lsun\ our mass-loss rates agree well with the \citeauthor{2000A&A...362..295V}
prescription -- which has been extensively scrutinized \citep[see e.g.][]
{2004A&A...415..349R} --
we focus the comparison of empirical and predicted mass-loss rates 
on lower luminosity stars, for which a `weak-wind problem' has been identified.

\subsubsection{Terminal velocities}
\label{sec:Observationtv}
Several studies have been devoted to measuring the terminal velocities of early-type stars.
Summarizing the work by \cite{1989ApJS...69..527H,1990ApJ...361..607P,1995ApJ...455..269L,
1997MNRAS.284..265H,1996A&A...305..171P}, and \cite{1999A&A...350..970K}, 
\cite{2000ARA&A..38..613K} 
derive that the average value of empirically determined terminal 
velocities for stars hotter than 21\,000\,K is $\vinf = 2.65\,\vesc$. The quoted accuracy 
of this mean value is roughly 20 percent. The \vinf\ values are ``measured'' from 
the maximum blue-shifted absorption $v_{\rm max}$ in resonance lines of ions such as 
C\,{\sc iv}, N\,{\sc v} and Si\,{\sc iv}, located in the ultraviolet part of the spectrum. These 
measurements are prone to systematic uncertainties, that have been extensive discussed in the
literature (see for instance the above references). They may work in both directions.
Effects that may cause the terminal velocity to be higher than $v_{\rm max}$ are 
measurements from lines that are not saturated in the outer wind (where for all practical
purposes \vinf\ is reached) or from ions that recombine in the outer wind. The former may be
expected for stars with weak winds, the latter is more likely to occur in very dense winds.
Effects that may cause \vinf\ to be smaller than $v_{\rm max}$ may be the presence of
turbulence in the outflow or the presence of strong atmospheric absorption at wavelengths
slightly bluer than the wavelength corresponding to the terminal velocity, mistakenly
contributed to absorption in the resonance line.
Given the possible occurrence of these systematic effects, the uncertainty in
the terminal velocity may be 10-15 percent for supergiants, and substantially larger than 20 percent for dwarfs.
The error in the ratio $\vinf/\vesc$ also includes 
uncertainties in \vesc. The largest contribution to this error comes from uncertainties
in the stellar masses, that have been derived from a comparison to tracks of
stellar evolution. It seems realistic to adopt a 30-40 percent uncertainty in
the empirical values of $\vinf/\vesc$ rather than the 20 percent; quoted at the beginning of this section.

Although \cite{2000ARA&A..38..613K} (and also \citeauthor{1995ApJ...455..269L} 
\citeyear{1995ApJ...455..269L}) conclude that the ratio $\vinf/\vesc$ is constant
for O-type stars, the results of \citet{1989ApJS...69..527H} show this ratio to decrease with
temperature, from about 3.5 at 31\,500\,K to about 2.4 at 43\,500\,K. A luminosity class dependence
of $\vinf/\vesc$ has to our knowledge not yet been reported. 

\begin{figure}
  \centering 
     \resizebox{8cm}{!}{
     \includegraphics[width=0.70\textwidth,angle=0]{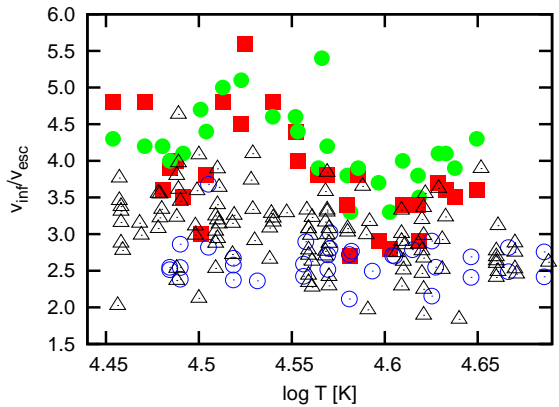}
      } 
  \caption{Predictions of the ratio of terminal velocity over effective escape velocity at the surface
              for both the best-$\beta$ (in red) and hydrodynamic method (in green) as a function of effective temperature.
              The blue circles denote the data from \cite{1995ApJ...455..269L} and the black triangles show the empirical 
	      values from \cite{1989ApJS...69..527H} in which \vinf\ is determined from the ultraviolet P-Cygni profiles.
 	        }
  \label{fig:observationalvelocities}
\end{figure}

Figure~\ref{fig:observationalvelocities} shows our predictions of $\vinf/\vesc$ plotted
against temperature. Different symbol types denote best-$\beta$,
hydrodynamic results and empirical values from \cite{1989ApJS...69..527H} and \cite{1995ApJ...455..269L}.
As discussed in Sections~\ref{sec:bestbeta} and~\ref{sec:ns} our predictions 
of \vinf\ have individual random error bars of 10 to 20 percent

In all cases, our predictions result in terminal velocities that are larger than observed.
For the main sequence O3-O6 stars the mean predicted ratio is  3.1 for the best-$\beta$
method and 3.6 for the hydrodynamical method. This is 17\% and 36\% higher than the observed
mean value of 2.65.
For giants the discrepancies are respectively 30\% and 44\%. For the supergiants
the largest discrepancies are found. Using all O stars, the best-$\beta$ method 
over-predicts the ratio by 54\%. The hydrodynamical method yields values that are on
average 60\% higher. The discrepancy between theory and observations thus seems
to increase from dwarfs to giants to supergiants. Given the uncertainties, the 
over-prediction for the dwarfs may not be significant.

The predictions show a tentative trend of a decreasing $\vinf / \vesc$ with temperature.
As pointed out, recent empirical studies do not recover this behavior.
Interestingly, this type of trend appears to be visible in the study by \citet{1989ApJS...69..527H}.
Their trend is plotted in Fig.~\ref{fig:observationalvelocities}, featuring a slope that is comparable 
to the slope of our predictions.
However, given the uncertainties in the current empirical estimates of \vinf, we do not feel that this
can be applied to (further) scrutinize the theory.

Larger predicted terminal velocities are also reported by \cite{1995ApJ...455..269L}.
In their sample, dominated by supergiants, a comparison to CAK models yields
over-predictions by about 33 percent, so slightly less compared to what we find.
The reason for the over-predicted \vinf\ values is unclear. 
Possible explanations (for part of the problem) include, 
{\em i)} overestimated corrections for the effect of turbulence (see above),
{\em ii)} a clumped and porous outer wind, hampering the acceleration of the flow
in this part of the outflow from reaching as high a terminal velocity as predicted here
and in (modified) CAK
\cite[see][]{2011A&A...526A..32M}, or {\em iii)}
a systematic over-estimate of stellar masses. 
A systematic discrepancy between
masses of galactic stars derived from comparing their positions in the Hertzsprung-Russell diagram to
evolutionary tracks and masses calculated from the spectroscopically determined gravity 
was reported by e.g. \cite{1992A&A...261..209H} 
\citep[but see][]{2010A&A...524A..98W}. Improvements in both the model
atmospheres and fitting procedure seem to have reduced, but possibly 
not yet eliminated, the size of this discrepancy \citep{2004A&A...415..349R,
2005A&A...441..711M}.

Unfortunately, progress in resolving the differences in predicted and empirical
 $\vinf/\vesc$ ratios quite strongly depends on our knowledge of stellar masses.
 Hopefully, detailed studies of very large populations, such as the VLT-FLAMES Survey of massive
 stars \citep{2005A&A...437..467E} and the VLT-FLAMES Tarantula Survey
 \citep{2011A&A...530A.108E} may help resolve this issue.

\subsubsection{Mass loss rates: the weak-wind problem}
\label{sec:weakwind}

\begin{figure}
  \centering 
     \resizebox{9cm}{!}{
      \includegraphics[width=0.70\textwidth,angle=0]{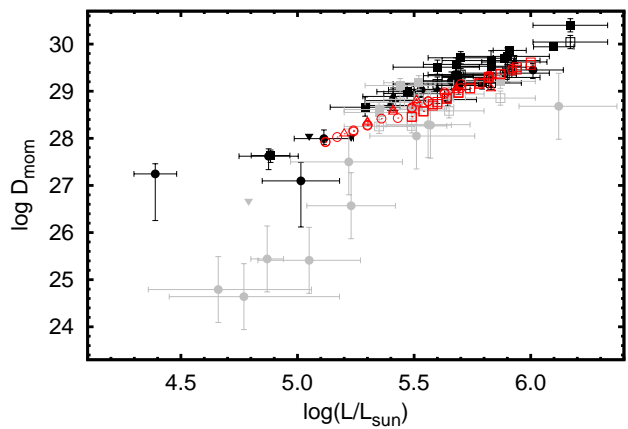}
}
  \caption{The modified wind momentum $D_{\rm mom} = \sqrt(\rstar/\rsun) \mdot \vinf$ as a function of stellar luminosity using the data 
           of \cite{2007A&A...473..603M}. Black symbols refer to mass-loss estimates based on the fitting of the H$\alpha$-profile; grey 
	   symbols are mass-loss estimates that rely strongly on ultraviolet resonance lines. Note that the H$\alpha$ estimates at $L < 10^{5.2}\,\lsun$ are upper limits.
	   A steep jump--of about 2 dex--can be seen at a luminosity of $10^{5.2}$\,\lsun. 
           The red dots are our predictions for $D_{\rm mom}$. The squares denote the supergiants,
	   the circles the giants and the triangles the main-sequence stars. Below  $10^{5.2}$\,\lsun\
	   we do not find wind solutions for dwarf stars. At higher luminosity our predictions are close to the observed values. 
           We therefore interpret the origin of the weak-wind
	   problem in dwarf stars to be connected to a lack of line driving for objects less bright than about $10^{5.2}$\,\lsun\ .}
  \label{fig:modified}
\end{figure}

Relatively recent, analysis of appreciable samples of galactic stars using sophisticated model atmospheres
has revealed a mismatch, possibly as high as a factor of 100, between empirically derived mass-loss rates
and theoretical predictions for stars less luminous than about $10^{5.2}$\,
\lsun\ \citep[see e.g.][and Fig.~\ref{fig:modified}]{2005A&A...441..735M,
2007A&A...473..603M,2009A&A...498..837M}. As mass loss scales with some power of the luminosity, this
problem occurs below a critical mass-flux and is termed the `weak-wind' problem 
\citep[for a recent review, see][]{2008A&ARv..16..209P}. Proposed explanations address 
deficiencies in determining the empirical mass-loss rates as well as in mass-loss predictions.
Regarding empirical \mdot\ determinations it should be realized that only UV resonance lines can be
used as a diagnostic in the weak-wind regime, whilst in the (lets call it the) strong wind regime H$\alpha$ 
and, in the Galactic case, radio-fluxes may also be used. The ions that produce the UV resonance
profiles, such as C\,{\sc iv}, N\,{\sc v} and Si\,{\sc iv}, often represent minor ionization species. 
The ionization continua of these species border the soft X-ray regime and therefore wind material
may be susceptible to (non-thermal) processes producing soft X-ray emission, such as shocks or
magnetic mechanisms \citep{2005A&A...441..735M}.

From a theoretical viewpoint, potential causes of the weak-wind problem include the
decoupling of the major driving ions (the metals) from the bulk of the plasma at low densities, when 
Coulomb coupling fails, and the subsequent ionic runaway 
\citep{1992A&A...262..515S,1995A&A...301..823B,1996A&A...309..867B,2000A&A...359..983K,
2002ApJ...568..965O,2003A&A...402..713K};  the shadowing of wind-driving lines by photospheric
lines \citep{1996A&A...309..867B}, and the neglect of curvature terms in the velocity field
\citep{1998ASPC..131..245P,1999ApJ...510..355O}.

The results presented in this paper point to a cause for the weak winds related to the 
predictions of mass loss. This potential cause was quantitatively explored by \cite{2010A&A...512A..33L},
who pointed out that the global dynamical constraint imposed by
\cite{2000A&A...362..295V} and recapped in Sec. \ref{sec:mcwind} (notably Eq. \ref{eq:mcdeltal}) need not 
guarantee that the derived mass-loss rates are consistent with stationary trans-sonic flows. Here we have 
shown that although this assumption by \cite{2000A&A...362..295V} is allowed for stars with luminosities above
$10^{5.2}$\,\lsun, it is not for lower luminosities. 
This luminosity limit for galactic O stars agrees with the empirical limit at $\sim 10^{5.2}$\,\lsun\ to within
0.1 dex in $\log L$.
The physical 
cause of the different \mdot\ regimes (weak and strong winds) is a lack of line acceleration at 
the base of the wind. In main sequence stars, a contribution of Fe\,{\sc v} lines is present
in O6 stars and is missing in the (lower luminosity and cooler) O6.5 stars, where Fe\,{\sc iv}
is more dominant (see Sect.~\ref{sec:lateO}).
The importance of the Fe\,{\sc v}/{\sc iv} ionization balance has been pointed out by
\cite{2010A&A...512A..33L} and is confirmed by our results. 

In one fundamental aspect our results differ from that of \cite{2010A&A...512A..33L}. For the sample
of low luminosity stars (i.e. less than $10^{5.2}$\,\lsun) investigated by \citet{2009A&A...498..837M}, 
\cite{2010A&A...512A..33L} predicts mass-loss rates that are about 1.4 dex lower than anticipated by 
\cite{2000A&A...362..295V}. The hydrodynamical method presented in this paper, which identifies
the ability of the star to drive an outflow with a balance of the line force and the gravitational force 
at the sonic point, predicts that these stars do not have a wind at all.
As it is clear form the presence of the shape of UV resonance lines that these 
stars do have stellar winds (with average mass loss rates that are 0.8 dex lower than 
\citeauthor{2010A&A...512A..33L}'s predictions), our result suggests that either some
other mechanism is driving the wind or is supplementing the line acceleration at the base
of the wind. Which force (or forces) 
counterbalances gravity remains to be identified, but perhaps magnetic 
pressure, effects of turbulence, and/or pulsations may play a role. 

We do point out that once material is accelerated, there is sufficient opacity available to further accelerate it to larger velocities.
Interestingly, \cite{2005A&A...441..735M} report that for their sample of galactic weak-wind
objects the average value of $\vinf/\vesc$ is rather close to unity, and not 2.65. Although
they concede that given the low wind densities their \vinf\ values may be lower limits,
they point to a mechanism of X-ray heating proposed by \cite{1994MNRAS.266..917D}
that may perhaps explain these results. In the outer atmospheres of weak winds the
cooling times can become quite long, such that heating of the material in for instance
shocks may warm up the medium and strongly modify the ionization structure, in effect 
canceling the line force.

Modified CAK theory does not predict the weak-wind discontinuity. In this theory the adopted values for $k$ and
$\alpha$ are based on the input stellar spectrum, while the dilution and excitation/ionization changes throughout the wind
are described by a fixed $\delta$. Therefore, no self-consistent feedback between the wind properties and the line acceleration is accounted 
for. We note that in the predictions by \cite{2001A&A...375..161P} a change of slope of the modified wind momentum luminosity relation 
can be seen at a luminosity of about $10^{5.2}$\,\lsun. It is tentative to suggest that if \citeauthor{2001A&A...375..161P} would have implemented
the iterative procedure that we use, they might have identified a weak-wind regime on theoretical grounds.

\section{Conclusions}
\label{sec:Conclusions}

We have presented new mass-loss rates and terminal wind velocities for a grid of massive $\textrm{O}$-type 
stars, improving the treatment of physics in the Monte Carlo method by 
\citet{1985ApJ...288..679A} and \citet{1997ApJ...477..792D} to predict wind properties of early-type stars.
Two new types of solutions have been discussed.
First, building on the work of \citet{2008A&A...492..493M}, we present so-called best-$\beta$ solutions in which
one still assumes a $\beta$-type velocity law (see Eq.~\ref{eq:betalaw}) in the wind, but in which the terminal 
velocity and $\beta$ are no longer adopted but constrained by requiring that they best fit the line force
(distribution).
Second, we abandon the $\beta$-type velocity structure and introduce numerical solutions 
of the wind stratification. Our main conclusions are:

\begin{enumerate}

\item For stars more luminous than $10^{5.2}$\,\lsun\, the best-$\beta$ and hydrodynamical method
      yield $\beta$ and $\vinf$ results in agreement with each other (within 5-20 percent), 
      whilst the mass-loss rates agree within a factor of 2.

\item Furthermore, both methods are in very good agreement with the mass-loss prescription
        by \cite{2000A&A...362..295V} using our terminal velocities in their recipe. This implies that the
        main assumption entering the method on which the \citeauthor{2000A&A...362..295V}
        results are based -- i.e. that the momentum equation is not solved explicitly -- {\em is not
        compromising their predicted \mdot\ in this luminosity range}. Terminal velocity is an input parameter 
	to the \citeauthor{2000A&A...362..295V} recipe. If we apply the canonical value $\vinf = 2.6 \varv_{\rm esc}$, 
	the discrepancy between our mass-loss rates and their mass-loss rates 
	is of the order of $0.2 \textrm{, although occasionally }0.6 \textrm{ dex}$.
        
\item At luminosities $\la 10^{5.2}$\,\lsun\  our hydrodynamical method fails to produce an outflow
        because of a lack of line driving at the base of the wind. This critical
        luminosity coincides with the onset of the `weak-wind problem'. 
                
\item For O dwarfs the above luminosity criterion translates to a boundary between starting and 
        failing to start a wind at O6/O6.5.
        The direct cause of the failure to start a wind in O6.5 V stars is the lower luminosity and
        the lack of Fe\,{\sc v} lines at the base of the wind compared to O6 V stars. 
        
\item The fact that our hydrodynamical method fails to drive a wind at $L \la 10^{5.2}$ \lsun\ may imply that
        some other mechanism is driving the weak winds or is
        supplementing the line acceleration at the base of the wind to help drive gas and initiate the wind.
        
\item For stars more luminous than $10^{5.2}$\,\lsun\ we predict, using the best-$\beta$ and hydrodynamical method,
        terminal velocities that are typically 35 and 45 percent higher than observed. Such
        over-predictions are similar to what is seen in MCAK-theory \citep{1995ApJ...455..269L}.
        
\item We predict beta values in the range 0.85 to 1.05, with a trend that supergiants have slightly
      higher $\beta$ values than dwarfs. 
      This range of $\beta$ values agrees very well with empirical results by \citet{2003ApJ...586..996M}.

\end{enumerate}

\begin{acknowledgements}
We would like to thank the referee Achim Feldmeier for constructive comments.
\end{acknowledgements}

\bibliographystyle{aa}
\bibliography{references.bib}

\begin{thebibliography}{66}
\expandafter\ifx\csname natexlab\endcsname\relax\def\natexlab#1{#1}\fi

\bibitem[{{Abbott}(1982)}]{1982ApJ...259..282A}
{Abbott}, D.~C. 1982, \apj, 259, 282

\bibitem[{{Abbott} \& {Lucy}(1985)}]{1985ApJ...288..679A}
{Abbott}, D.~C. \& {Lucy}, L.~B. 1985, \apj, 288, 679

\bibitem[{{Anders} \& {Grevesse}(1989)}]{1989GeCoA..53..197A}
{Anders}, E. \& {Grevesse}, N. 1989, \gca, 53, 197

\bibitem[{{Babel}(1995)}]{1995A&A...301..823B}
{Babel}, J. 1995, \aap, 301, 823

\bibitem[{{Babel}(1996)}]{1996A&A...309..867B}
{Babel}, J. 1996, \aap, 309, 867

\bibitem[{{Brott} {et~al.}(2009){Brott}, {Hunter}, {de Koter}, {Langer},
  {Lennon}, \& {Dufton}}]{2009CoAst.158...55B}
{Brott}, I., {Hunter}, I., {de Koter}, A., {et~al.} 2009, Communications in
  Asteroseismology, 158, 55

\bibitem[{{Castor} {et~al.}(1975){Castor}, {Abbott}, \&
  {Klein}}]{1975ApJ...195..157C}
{Castor}, J.~I., {Abbott}, D.~C., \& {Klein}, R.~I. 1975, \apj, 195, 157

\bibitem[{{Chiosi} \& {Maeder}(1986)}]{1986ARA&A..24..329C}
{Chiosi}, C. \& {Maeder}, A. 1986, \araa, 24, 329

\bibitem[{{de Koter} {et~al.}(1997){de Koter}, {Heap}, \&
  {Hubeny}}]{1997ApJ...477..792D}
{de Koter}, A., {Heap}, S.~R., \& {Hubeny}, I. 1997, \apj, 477, 792

\bibitem[{{de Koter} {et~al.}(1993){de Koter}, {Schmutz}, \&
  {Lamers}}]{1993A&A...277..561D}
{de Koter}, A., {Schmutz}, W., \& {Lamers}, H.~J.~G.~L.~M. 1993, \aap, 277, 561

\bibitem[{{Drew} {et~al.}(1994){Drew}, {Hoare}, \&
  {Denby}}]{1994MNRAS.266..917D}
{Drew}, J.~E., {Hoare}, M.~G., \& {Denby}, M. 1994, \mnras, 266, 917

\bibitem[{{Eldridge} \& {Vink}(2006)}]{2006A&A...452..295E}
{Eldridge}, J.~J. \& {Vink}, J.~S. 2006, \aap, 452, 295

\bibitem[{{Evans} {et~al.}(2005){Evans}, {Smartt}, {Lee}, {Lennon}, {Kaufer},
  {Dufton}, {Trundle}, {Herrero}, {Sim{\'o}n-D{\'{\i}}az}, {de Koter},
  {Hamann}, {Hendry}, {Hunter}, {Irwin}, {Korn}, {Kudritzki}, {Langer},
  {Mokiem}, {Najarro}, {Pauldrach}, {Przybilla}, {Puls}, {Ryans}, {Urbaneja},
  {Venn}, \& {Villamariz}}]{2005A&A...437..467E}
{Evans}, C.~J., {Smartt}, S.~J., {Lee}, J., {et~al.} 2005, \aap, 437, 467

\bibitem[{{Evans} {et~al.}(2011){Evans}, {Taylor}, {H{\'e}nault-Brunet},
  {Sana}, {de Koter}, {Sim{\'o}n-D{\'{\i}}az}, {Carraro}, {Bagnoli}, {Bastian},
  {Bestenlehner}, {Bonanos}, {Bressert}, {Brott}, {Campbell}, {Cantiello},
  {Clark}, {Costa}, {Crowther}, {de Mink}, {Doran}, {Dufton}, {Dunstall},
  {Friedrich}, {Garcia}, {Gieles}, {Gr{\"a}fener}, {Herrero}, {Howarth},
  {Izzard}, {Langer}, {Lennon}, {Ma{\'{\i}}z Apell{\'a}niz}, {Markova},
  {Najarro}, {Puls}, {Ramirez}, {Sab{\'{\i}}n-Sanjuli{\'a}n}, {Smartt},
  {Stroud}, {van Loon}, {Vink}, \& {Walborn}}]{2011A&A...530A.108E}
{Evans}, C.~J., {Taylor}, W.~D., {H{\'e}nault-Brunet}, V., {et~al.} 2011, \aap,
  530, A108+

\bibitem[{{Feldmeier} \& {Shlosman}(2000)}]{2000ApJ...532L.125F}
{Feldmeier}, A. \& {Shlosman}, I. 2000, \apjl, 532, L125

\bibitem[{{Herrero} {et~al.}(1992){Herrero}, {Kudritzki}, {Vilchez}, {Kunze},
  {Butler}, \& {Haser}}]{1992A&A...261..209H}
{Herrero}, A., {Kudritzki}, R.~P., {Vilchez}, J.~M., {et~al.} 1992, \aap, 261,
  209

\bibitem[{{Howarth} \& {Prinja}(1989)}]{1989ApJS...69..527H}
{Howarth}, I.~D. \& {Prinja}, R.~K. 1989, \apjs, 69, 527

\bibitem[{{Howarth} {et~al.}(1997){Howarth}, {Siebert}, {Hussain}, \&
  {Prinja}}]{1997MNRAS.284..265H}
{Howarth}, I.~D., {Siebert}, K.~W., {Hussain}, G.~A.~J., \& {Prinja}, R.~K.
  1997, \mnras, 284, 265

\bibitem[{{Krti{\v c}ka} \& {Kub{\'a}t}(2000)}]{2000A&A...359..983K}
{Krti{\v c}ka}, J. \& {Kub{\'a}t}, J. 2000, \aap, 359, 983

\bibitem[{{Krti{\v c}ka} \& {Kub{\'a}t}(2004)}]{2004A&A...417.1003K}
{Krti{\v c}ka}, J. \& {Kub{\'a}t}, J. 2004, \aap, 417, 1003

\bibitem[{{Krti{\v c}ka} {et~al.}(2003){Krti{\v c}ka}, {Owocki}, {Kub{\'a}t},
  {Galloway}, \& {Brown}}]{2003A&A...402..713K}
{Krti{\v c}ka}, J., {Owocki}, S.~P., {Kub{\'a}t}, J., {Galloway}, R.~K., \&
  {Brown}, J.~C. 2003, \aap, 402, 713

\bibitem[{{Kudritzki} \& {Puls}(2000)}]{2000ARA&A..38..613K}
{Kudritzki}, R. \& {Puls}, J. 2000, \araa, 38, 613

\bibitem[{{Kudritzki}(2002)}]{2002ApJ...577..389K}
{Kudritzki}, R.~P. 2002, \apj, 577, 389

\bibitem[{{Kudritzki} {et~al.}(1999){Kudritzki}, {Puls}, {Lennon}, {Venn},
  {Reetz}, {Najarro}, {McCarthy}, \& {Herrero}}]{1999A&A...350..970K}
{Kudritzki}, R.~P., {Puls}, J., {Lennon}, D.~J., {et~al.} 1999, \aap, 350, 970

\bibitem[{{Lamers} \& {Cassinelli}(1999)}]{1999isw..book.....L}
{Lamers}, H.~J.~G.~L.~M. \& {Cassinelli}, J.~P. 1999, {Introduction to Stellar
  Winds} (Introduction to Stellar Winds, by Henny J.~G.~L.~M.~Lamers and Joseph
  P.~Cassinelli, pp.~452.~ISBN 0521593980.~Cambridge, UK: Cambridge University
  Press, June 1999.)

\bibitem[{{Lamers} {et~al.}(1995){Lamers}, {Snow}, \&
  {Lindholm}}]{1995ApJ...455..269L}
{Lamers}, H.~J.~G.~L.~M., {Snow}, T.~P., \& {Lindholm}, D.~M. 1995, \apj, 455,
  269

\bibitem[{{Langer}(1998)}]{1998A&A...329..551L}
{Langer}, N. 1998, \aap, 329, 551

\bibitem[{{Limongi} \& {Chieffi}(2006)}]{2006ApJ...647..483L}
{Limongi}, M. \& {Chieffi}, A. 2006, \apj, 647, 483

\bibitem[{{Lucy}(2010)}]{2010A&A...512A..33L}
{Lucy}, L.~B. 2010, \aap, 512, A33

\bibitem[{{Lucy} \& {Solomon}(1970)}]{1970ApJ...159..879L}
{Lucy}, L.~B. \& {Solomon}, P.~M. 1970, \apj, 159, 879

\bibitem[{{Maeder}(1981)}]{1981A&A....99...97M}
{Maeder}, A. 1981, \aap, 99, 97

\bibitem[{{Maeder} \& {Meynet}(2000)}]{2000A&A...361..159M}
{Maeder}, A. \& {Meynet}, G. 2000, \aap, 361, 159

\bibitem[{{Marcolino} {et~al.}(2009){Marcolino}, {Bouret}, {Martins},
  {Hillier}, {Lanz}, \& {Escolano}}]{2009A&A...498..837M}
{Marcolino}, W.~L.~F., {Bouret}, J., {Martins}, F., {et~al.} 2009, \aap, 498,
  837

\bibitem[{{Martins} {et~al.}(2005{\natexlab{a}}){Martins}, {Schaerer}, \&
  {Hillier}}]{2005A&A...436.1049M}
{Martins}, F., {Schaerer}, D., \& {Hillier}, D.~J. 2005{\natexlab{a}}, \aap,
  436, 1049

\bibitem[{{Martins} {et~al.}(2005{\natexlab{b}}){Martins}, {Schaerer},
  {Hillier}, {Meynadier}, {Heydari-Malayeri}, \&
  {Walborn}}]{2005A&A...441..735M}
{Martins}, F., {Schaerer}, D., {Hillier}, D.~J., {et~al.} 2005{\natexlab{b}},
  \aap, 441, 735

\bibitem[{{Massa} {et~al.}(2003){Massa}, {Fullerton}, {Sonneborn}, \&
  {Hutchings}}]{2003ApJ...586..996M}
{Massa}, D., {Fullerton}, A.~W., {Sonneborn}, G., \& {Hutchings}, J.~B. 2003,
  \apj, 586, 996

\bibitem[{{Meynet} \& {Maeder}(2003)}]{2003A&A...404..975M}
{Meynet}, G. \& {Maeder}, A. 2003, \aap, 404, 975

\bibitem[{{Mokiem} {et~al.}(2005){Mokiem}, {de Koter}, {Puls}, {Herrero},
  {Najarro}, \& {Villamariz}}]{2005A&A...441..711M}
{Mokiem}, M.~R., {de Koter}, A., {Puls}, J., {et~al.} 2005, \aap, 441, 711

\bibitem[{{Mokiem} {et~al.}(2007){Mokiem}, {de Koter}, {Vink}, {Puls}, {Evans},
  {Smartt}, {Crowther}, {Herrero}, {Langer}, {Lennon}, {Najarro}, \&
  {Villamariz}}]{2007A&A...473..603M}
{Mokiem}, M.~R., {de Koter}, A., {Vink}, J.~S., {et~al.} 2007, \aap, 473, 603

\bibitem[{{Muijres} {et~al.}(2011){Muijres}, {de Koter}, {Vink}, {Krti{\v
  c}ka}, {Kub{\'a}t}, \& {Langer}}]{2011A&A...526A..32M}
{Muijres}, L.~E., {de Koter}, A., {Vink}, J.~S., {et~al.} 2011, \aap, 526, A32+

\bibitem[{{M{\"u}ller}(2001)}]{2001mueller}
{M{\"u}ller}, P.~E. 2001, {Ph.D. Thesis, Univ. of Heidelberg, Germany,
  \texttt{www.ub.uni-heidelberg.de/archiv/1422} }

\bibitem[{{M{\"u}ller} \& {Vink}(2008)}]{2008A&A...492..493M}
{M{\"u}ller}, P.~E. \& {Vink}, J.~S. 2008, \aap, 492, 493

\bibitem[{{Owocki} \& {Puls}(1999)}]{1999ApJ...510..355O}
{Owocki}, S.~P. \& {Puls}, J. 1999, \apj, 510, 355

\bibitem[{{Owocki} \& {Puls}(2002)}]{2002ApJ...568..965O}
{Owocki}, S.~P. \& {Puls}, J. 2002, \apj, 568, 965

\bibitem[{{Palacios} {et~al.}(2005){Palacios}, {Arnould}, \&
  {Meynet}}]{2005A&A...443..243P}
{Palacios}, A., {Arnould}, M., \& {Meynet}, G. 2005, \aap, 443, 243

\bibitem[{{Pauldrach} {et~al.}(1986){Pauldrach}, {Puls}, \&
  {Kudritzki}}]{1986A&A...164...86P}
{Pauldrach}, A., {Puls}, J., \& {Kudritzki}, R.~P. 1986, \aap, 164, 86

\bibitem[{{Pauldrach} {et~al.}(2001){Pauldrach}, {Hoffmann}, \&
  {Lennon}}]{2001A&A...375..161P}
{Pauldrach}, A.~W.~A., {Hoffmann}, T.~L., \& {Lennon}, M. 2001, \aap, 375, 161

\bibitem[{{Press} {et~al.}(1992){Press}, {Teukolsky}, {Vetterling}, \&
  {Flannery}}]{1992nrfa.book.....P}
{Press}, W.~H., {Teukolsky}, S.~A., {Vetterling}, W.~T., \& {Flannery}, B.~P.
  1992, {Numerical recipes in FORTRAN. The art of scientific computing}, ed.
  {Press, W.~H., Teukolsky, S.~A., Vetterling, W.~T., \& Flannery, B.~P. }

\bibitem[{{Prinja} {et~al.}(1990){Prinja}, {Barlow}, \&
  {Howarth}}]{1990ApJ...361..607P}
{Prinja}, R.~K., {Barlow}, M.~J., \& {Howarth}, I.~D. 1990, \apj, 361, 607

\bibitem[{{Puls}(1987)}]{1987A&A...184..227P}
{Puls}, J. 1987, \aap, 184, 227

\bibitem[{{Puls} {et~al.}(1996){Puls}, {Kudritzki}, {Herrero}, {Pauldrach},
  {Haser}, {Lennon}, {Gabler}, {Voels}, {Vilchez}, {Wachter}, \&
  {Feldmeier}}]{1996A&A...305..171P}
{Puls}, J., {Kudritzki}, R., {Herrero}, A., {et~al.} 1996, \aap, 305, 171

\bibitem[{{Puls} {et~al.}(1998){Puls}, {Kudritzki}, {Santolaya-Rey}, {Herrero},
  {Owocki}, \& {McCarthy}}]{1998ASPC..131..245P}
{Puls}, J., {Kudritzki}, R., {Santolaya-Rey}, A.~E., {et~al.} 1998, in
  Astronomical Society of the Pacific Conference Series, Vol. 131, Properties
  of Hot Luminous Stars, ed. {I.~Howarth}, 245--

\bibitem[{{Puls} {et~al.}(2000){Puls}, {Springmann}, \&
  {Lennon}}]{2000A&AS..141...23P}
{Puls}, J., {Springmann}, U., \& {Lennon}, M. 2000, \aaps, 141, 23

\bibitem[{{Puls} {et~al.}(2008){Puls}, {Vink}, \&
  {Najarro}}]{2008A&ARv..16..209P}
{Puls}, J., {Vink}, J.~S., \& {Najarro}, F. 2008, \aapr, 16, 209

\bibitem[{{Repolust} {et~al.}(2004){Repolust}, {Puls}, \&
  {Herrero}}]{2004A&A...415..349R}
{Repolust}, T., {Puls}, J., \& {Herrero}, A. 2004, \aap, 415, 349

\bibitem[{{Schaerer} \& {de Koter}(1997)}]{1997A&A...322..598S}
{Schaerer}, D. \& {de Koter}, A. 1997, \aap, 322, 598

\bibitem[{{Sobolev}(1960)}]{1960mes..book.....S}
{Sobolev}, V.~V. 1960, {Moving envelopes of stars}, ed. V.~V. {Sobolev}

\bibitem[{{Springmann} \& {Pauldrach}(1992)}]{1992A&A...262..515S}
{Springmann}, U.~W.~E. \& {Pauldrach}, A.~W.~A. 1992, \aap, 262, 515

\bibitem[{{Vink}(2000)}]{2000PhDT........45V}
{Vink}, J.~S. 2000, PhD thesis, Universiteit Utrecht

\bibitem[{{Vink} {et~al.}(2010){Vink}, {Brott}, {Gr{\"a}fener}, {Langer}, {de
  Koter}, \& {Lennon}}]{2010A&A...512L...7V}
{Vink}, J.~S., {Brott}, I., {Gr{\"a}fener}, G., {et~al.} 2010, \aap, 512, L7

\bibitem[{{Vink} \& {de Koter}(2002)}]{2002A&A...393..543V}
{Vink}, J.~S. \& {de Koter}, A. 2002, \aap, 393, 543

\bibitem[{{Vink} \& {de Koter}(2005)}]{2005A&A...442..587V}
{Vink}, J.~S. \& {de Koter}, A. 2005, \aap, 442, 587

\bibitem[{{Vink} {et~al.}(1999){Vink}, {de Koter}, \&
  {Lamers}}]{1999A&A...350..181V}
{Vink}, J.~S., {de Koter}, A., \& {Lamers}, H.~J.~G.~L.~M. 1999, \aap, 350, 181

\bibitem[{{Vink} {et~al.}(2000){Vink}, {de Koter}, \&
  {Lamers}}]{2000A&A...362..295V}
{Vink}, J.~S., {de Koter}, A., \& {Lamers}, H.~J.~G.~L.~M. 2000, \aap, 362, 295

\bibitem[{{Vink} {et~al.}(2001){Vink}, {de Koter}, \&
  {Lamers}}]{2001A&A...369..574V}
{Vink}, J.~S., {de Koter}, A., \& {Lamers}, H.~J.~G.~L.~M. 2001, \aap, 369, 574

\bibitem[{{Weidner} \& {Vink}(2010)}]{2010A&A...524A..98W}
{Weidner}, C. \& {Vink}, J.~S. 2010, \aap, 524, A98+

\end{thebibliography}

\end{document}